\documentclass[a4paper,11pt]{article}

\usepackage{JPA-MT}
                     
\usepackage[english]{babel}
\usepackage[T1]{fontenc} 
\usepackage{newtxtext, newtxmath}

\usepackage{amssymb}
\usepackage{amsmath}
\usepackage{nicefrac}
\usepackage{nccmath}
\usepackage{amsfonts}                   
\usepackage{mathtools}
\usepackage{bbold}
\usepackage{dsfont}
\usepackage{marvosym}
\usepackage{physics}
\usepackage{listings}
\usepackage{relsize}
\usepackage{gauss}
\usepackage{empheq}

\usepackage{enumitem}
\usepackage{pdfpages}
\usepackage{graphics}
\usepackage{graphicx}
\usepackage{color}  
\usepackage{rotating}
\usepackage{tikz}
\usepackage{xcolor}
\usepackage{pstricks}
\usepackage{transparent}
\usepackage{soul}
\usepackage{etoolbox}

\usepackage[english]{babel}
\usepackage{scrextend}
\usepackage{multirow}
\usepackage[draft]{todonotes}
\usepackage{dirtytalk}
\usepackage[style=english]{csquotes}
\usepackage{tocloft}
\usepackage{ocg-p}
\usepackage{ocgx}
\usepackage{floatrow}
\usepackage{sidecap}
\usepackage{comment}
\usepackage[numbers]{natbib}
\usepackage{epigraph}

\usepackage[bottom]{footmisc}
\usepackage{slashed}
\usepackage{multicol}
\usepackage{fancyhdr}
\usepackage{verbatim}
\newcounter{eg}                         \newtheorem{eg}{Example}[section]        
\def\beg{\begin{eg}\rm}                 \def\eeg{\hfill\sq\end{eg}}

\DeclareMathOperator{\0}{\mathbb{0}}
\DeclareMathOperator{\1}{\mathbb{1}}
\DeclareMathOperator{\diag}{diag}

\newcommand{\Var}{\operatorname{Var}}

\makeatletter
\DeclareRobustCommand{\cev}[1]{%
  {\mathpalette\do@cev{#1}}%
}
\newcommand{\do@cev}[2]{%
  \vbox{\offinterlineskip
    \sbox\z@{$\m@th#1 x$}%
    \ialign{##\cr
      \hidewidth\reflectbox{$\m@th#1\vec{}\mkern4mu$}\hidewidth\cr
      \noalign{\kern-\ht\z@}
      $\m@th#1#2$\cr
    }%
  }%
}
\makeatother

\def\lr#1{\left(#1\right)}
\def\avg#1{\left\langle #1 \right\rangle}
\def\K{\mathcal K}

\usepackage{url}
\usepackage{xurl}
\PassOptionsToPackage{hyphens}{url}
\usepackage{hyperref}
\hypersetup{
    colorlinks=true,
    citecolor=[rgb]{0,0.334,0.666}, 
    linkcolor=[rgb]{0,0.334,0.666},
    urlcolor=[rgb]{0,0.334,0.666},
}

\usepackage{amsfonts} 
\usepackage{soul}
\usepackage{xcolor}

\def\C{\mathcal{C}}

\def\lr#1{\left(#1\right)}
\def\slr#1{\left[#1\right]}

\definecolor{newBlue}{RGB}{94,129,181}
\definecolor{newRed}{RGB}{197,110,26}
\definecolor{newGreen}{RGB}{143,177,49}

\title{\boldmath Pinpointing Triple Point of Noncommutative Matrix Model with Curvature}

\author[a]{D. Prekrat,
}
\author[b]{B. Bukor,}
\author[b]{and J. Tekel}

\affiliation[a]{University of Belgrade -- Faculty of Pharmacy,\\ Vojvode Stepe 450, Belgrade, Serbia}
\affiliation[b]{Department of Theoretical Physics, Faculty of Mathematics, Physics and Informatics, \\
Comenius University in Bratislava, \\
Mlynsk\'a dolina, 842 48, Bratislava, Slovakia}

\emailAdd{dragan.prekrat@pharmacy.bg.ac.rs, benedek.bukor@fmph.uniba.sk, juraj.tekel@fmph.uniba.sk}

\abstract{
We study a Hermitian matrix model with a quartic potential,
modified by a curvature term $\mathrm{tr}(R\Phi^2)$,
where $R$ is a fixed external matrix.
Inspired by the truncated Heisenberg algebra formulation
of the Grosse--Wulkenhaar model,
this term breaks unitary invariance and,
through perturbative expansion,
induces an effective multitrace matrix model.
We analyze the resulting action both analytically and numerically,
including Hamiltonian Monte Carlo simulations,
focusing on two features closely tied to renormalizability:
the shift of the triple point and the suppression
of the noncommutative striped phase.
Our findings show that the curvature term drives
the phase structure toward renormalizable behavior
by removing the striped phase in the large-$N$ limit,
while also unexpectedly revealing a possible novel
multi-cut phase observed at the level of finite matrix size.
}

\begin{document} 
\maketitle
\flushbottom

\section{Introduction}

One promising approach to reconciling the seemingly distinct realms of gravity and quantum theory is to modify the short-distance structure of spacetime \cite{Hossenfelder:2012jw}. In particular, this can be achieved by introducing coordinate noncommutativity (NC), initially proposed to incorporate both spacetime symmetries and high-energy cutoffs in quantum field theory \cite{PhysRev.71.38}. 
In this context, recent hints of a preferred direction in the observable
universe \cite{10.1093/mnras/staf292}, possibly associated with cosmic
rotation, are noteworthy, as such anisotropies are among the hallmarks of NC coordinates. At the same time, it is well established that quantum systems in magnetic fields can be described using effective NC coordinates \cite{Karabali:2002im,Karabali:2006eg}, suggesting additional NC effects may
also arise in the presence of magnetic fields on cosmological scales \cite{Han:2017htd}. Regardless of its possible origin, matrix models provide an ideal framework for studying NC physics due to intrinsic noncommutativity of matrix multiplication and well-defined path integrals. 

Matrix models also have extensive applications across various fields, including biophysics, solid-state physics, optics, nuclear physics, and quantum gravity \cite{Eynard:2015aea,Akemann:2011csh,PhysRevLett.94.168103,Beenakker:2014zza,Weidenmuller:2008vb,Guhr:1997ve,Loll:2019rdj}. They can serve as a means of regularizing quantum field theories \cite{Bietenholz:2014sza} and are conjectured to describe fundamental physical laws. In some models, matrix elements function as fields on spacetime, while in others, spacetime itself is absent as an explicit concept. Intriguingly, in certain parameter regimes, spacetime may emerge dynamically within these models \cite{Steinacker:2011ix}.

A major challenge in NC models is the phenomenon of UV/IR mixing \cite{SZABO2003207}, which entangles high- and low-energy scales, thereby complicating renormalization. This mixing disrupts the separation of energy scales—a crucial element of effective field theory \cite{Rosaler:2021quv}.

Over the past two decades, the Grosse--Wulkenhaar (GW) model \cite{Grosse:2003nw, Grosse:2004yu} has successfully addressed the renormalization issues of NC field theories by introducing an additional term in the action, which can be interpreted as arising\footnote{A similar term appears in the non-associative Snyder–de Sitter space \cite{Franchino-Vinas:2019nqy,Franchino-Vinas:2021bcl}, arising from the expansion of the kinetic term.} from curvature of the background NC space \cite{Buric:2009ss,Prekrat:2025ozs}. We have recently proposed that the nice behavior of the GW model is closely tied to its phase structure, particularly through the suppression of the NC striped phase \cite{Prekrat:2021uos, Prekrat:2022sir}. Our results in this regard were obtained using a matrix formulation of the GW model. Further recent exploration of phase structure of different NC matrix models can be found in \cite{Gubser:2000cd,Tekel:2015uza,Chen:2017kfj,Kovacik:2018thy,Han:2019wue,Kovacik:2020cod,Ydri:2020efr,Ydri:2021cam,Prekrat:2023thesis,Kovacik:2023zab,Bukor_2025} and, in particular, some models studying Dirac ensembles \cite{Barrett:2015foa, Khalkhali:2020djp, Hessam:2021byc, Hessam:2022gaw, Khalkhali:2023onm}.

The immediate motivation for this work was to derive an analytical prediction for the location of the triple point in the GW model using a sequence of perturbative approximations. To simplify the analysis, we considered a model without the kinetic term and used the turning points of the resulting approximate transition lines as proxies for the true triple point.

Despite the fact that such approximate transition lines generically diverge within the perturbative regime—a behavior also observed in other multitrace matrix models \cite{Subjakova:2020gdh}—we hoped that their turning points would converge toward the true nonperturbative position of the triple point.

This expectation was based on two observations. First, the turning point of the fourth-order approximation to the transition line lies closer to the expected location of the true triple point than the turning point of the exact transition line computed for the second-order effective action \cite{Prekrat:2022sir}. Second, as we will show in Section~6, successive perturbative approximations to the second-order action transition line exhibit turning points that appear to converge monotonically toward the exact result.

These observations led us to expect that a similar convergence pattern
might extend to higher-order approximations. However, our analysis up to sixth order in the curvature coupling indicates that such monotonic convergence does not occur—at least not at low perturbative orders.

This paper is organized as follows. In Section~2, we introduce the GW
model and present its matrix formulation, emphasizing the role of the
curvature term. Section~3 discusses the physical significance of phase
transitions in the context of renormalization and highlights the
connection between the triple point shift and the suppression of the
noncommutative striped phase. In Section~4, we derive the effective
action through a perturbative expansion and obtain analytical
expressions for the phase transition lines. Section~5 reviews the
matrix-model techniques used to analyze eigenvalue distributions and
the free energy. Section~6 focuses on the phase diagrams of truncated
multitrace submodels, while Section~7 presents a detailed comparison
with Hamiltonian Monte Carlo (HMC) simulations of the matrix GW model
without the kinetic term. Finally, in Section~8, we summarize our
findings and outline directions for future work.

\section{GW Model \& Matrix Action}

We begin by introducing our model and its underlying NC space.
The starting point is the two-dimensional GW-model \cite{Grosse:2003nw}

\begin{multline}
S_\text{\tiny GW} = \int \!\! \mathrm{d}^2x \,
\Bigg(
\frac{1}{2} \partial^\mu\phi \star \partial_\mu\phi
\;+\; \frac{m^2}{2}\phi\star\phi
\;+\; \frac{\lambda}{4!}\phi\star\phi\star\phi\star\phi
\;+\;
\\
\;+\; \frac{\Omega^2}{2}((\theta^{-1})_{\mu\rho}x^\rho\phi) \star ((\theta^{-1})^{\mu\sigma}x_\sigma\phi)
\Bigg) \, ,
\label{GW model}
\end{multline}
which lives on the Moyal plane equipped with a $\star$-product
\begin{equation}
    f \star g
    = 
    f\,e^{\nicefrac{\text{i}}{2}\,\cev{\partial}_\mu\theta^{\mu\nu}\vec{\partial}_\nu}\,g
\end{equation}
and with NC coordinates satisfying
\begin{equation}
    \comm{x^\mu}{x^\nu}_\star
    =
    \text{i}\theta^{\mu\nu}
    =
    \text{i}\theta\epsilon^{\mu\nu} \, .
\end{equation}
The first line in \eqref{GW model} correspond to the standard NC $\lambda\phi^4_\star$ model, which is non-renormalizable. However, the inclusion of the $\Omega$-term in the second line, renders the model superrenormalizable in two dimensions \cite{Wulkenhaar:habilitation2004}.

Applying the Weyl transform and promoting the field $\phi$ to an ${N \times N}$ Hermitian matrix $\Phi$ transforms the action \eqref{GW model} into a matrix model
\begin{equation}\label{GW matrix model}
S_\text{\tiny GW}^\text{\tiny M} = N\tr\!\lr{
\Phi\mathcal{K}\Phi
- g_r R\Phi^2
- g_2\Phi^2 + g_4\Phi^4
}
\end{equation}
on a background space spanned\footnote{Although the background space is three-dimensional (with two of the coordinates scaled as $\sqrt{N}X$, $\sqrt{N}Y$), we focus on rescaled coordinates and restrict to a subspace where the third coordinate is set to zero in a weak limit of infinite matrix size, which reproduces the GW-model. We also use a rescaled version of the curvature matrix $R$. For more details, see  \cite{Buric:2009ss}.} by NC coordinates
\begin{equation}
X = \frac{1}{\sqrt{2N}}
\begin{pmatrix} 
 & \scriptstyle+\sqrt{1} \\ 
 \scriptstyle+\sqrt{1} &  & \scriptstyle+\sqrt{2} \\ 
 & \scriptstyle+\sqrt{2} & &  \hspace{-25pt}\begin{rotate}{-5}{$\ddots$}\end{rotate} \\ 
 &  & \hspace{-25pt}\begin{rotate}{-5}{$\ddots$}\end{rotate} &  & \hspace{-16pt}\scriptstyle+\sqrt{N-1} \\ 
 &  & & \hspace{-15pt}\scriptstyle+\sqrt{N-1} &
\end{pmatrix} \, ,
\qquad
Y = \frac{\text{i}}{\sqrt{2N}}
\begin{pmatrix} 
 & \scriptstyle-\sqrt{1} \\ 
 \scriptstyle+\sqrt{1} &  & \scriptstyle-\sqrt{2} \\ 
 & \scriptstyle+\sqrt{2} & & \hspace{-25pt}\begin{rotate}{-5}{$\ddots$}\end{rotate} \\ 
 &   & \hspace{-25pt}\begin{rotate}{-5}{$\ddots$}\end{rotate} &  & \hspace{-16pt} \scriptstyle-\sqrt{N-1} \\ 
 &  &    & \hspace{-15pt}\scriptstyle+\sqrt{N-1} &
\end{pmatrix} \, .
\end{equation}
\\
\noindent
The price of introducing the finite matrix regularization of the NC coordinates is the modification of their commutation relations (${N \to \infty}$ limit restores the original ones) and curving of the initial Moyal space.  The curvature $R$ contains energy levels associated with the $\Omega$-term harmonic oscillator and is given by
\begin{equation}
R=\frac{15}{2N}-8\lr{X^{2}+Y^{2}} \stackrel{\scriptscriptstyle N \gg 1}{\approx} -\frac{16}{N} \diag\lr{1,2,\ldots,N}\, .
\end{equation}
The kinetic operator $\mathcal{K}$ is also quadratic in NC coordinates and is defined via double commutators 
\begin{equation}\label{kinetic}
\mathcal{K}\Phi = \comm{X}{\comm{X}{\Phi}} + \comm{Y}{\comm{Y}{\Phi}} \, .
\end{equation}

When constructing the model, we included the NC scale $\theta$ in the definitions of the matrix field and the couplings and set it to unity so that we could work with dimensionless quantities. Finally, we also introduce the unscaled\footnote{The unscaled parameters include a factor $N$ originating from the action \eqref{GW matrix model}.} versions of couplings
\begin{equation}
    G_2 = Ng_2 \, ,
    \qquad
    G_4 = Ng_4 \, ,
\end{equation}
as they will appear in the analytical results concerning renormalization.

\section{Phase Transitions \& Renormalization}

In this section, we discuss the significance of the phase transitions for the renormalization properties of the model under study.

\begin{figure}[t]
\centering 
\includegraphics[width=0.99\textwidth]{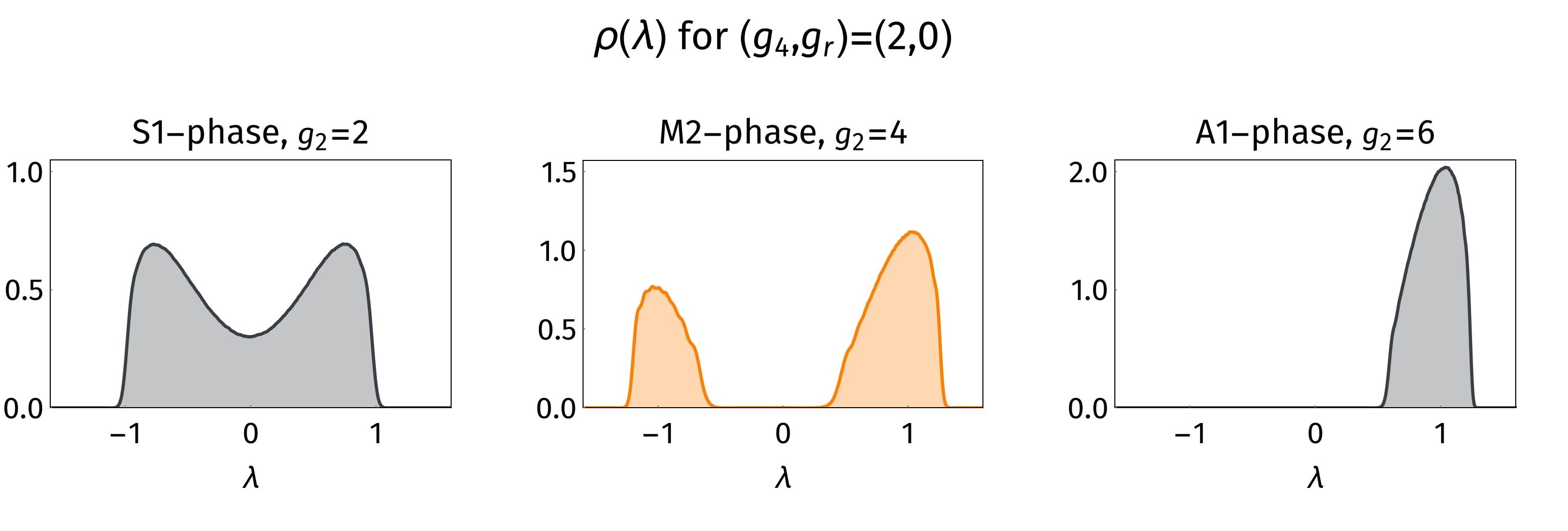}
\caption{
Eigenvalue distribution $\rho(\lambda)$ in the matrix GW model for representative phases (${N=24}$). The left and the right distributions correspond to the disordered and ordered vacua, respectively, both of which also appear in commutative models. The central orange distribution, the so-called striped (or matrix) vacuum, is unique to the noncommutative model and features eigenvalues of both signs. This phase breaks translational symmetry and leads to a spatial modulation of magnetization.}
\label{figure:01}
\end{figure}

The phase diagram of our model illustrates which vacuum solutions are realized across different parameter regimes. Applying the saddle point method to the matrix action \eqref{GW matrix model} yields the following equation of motion (EOM):
\begin{equation}\label{EOM}
2\mathcal{K}\Phi-g_r \lr{R \Phi + \Phi R}
+ 2\Phi\lr{2g_4\Phi^2 - g_2\!\1} = 0 \, .
\end{equation}
This equation has several relevant (approximate) solutions depending on the dominance of the kinetic term ($\mathcal{K}$), curvature term ($R$), or the pure potential\footnote{We consider the regime $g_2>0$, where spontaneous symmetry breaking occurs.} terms (involving $g_2$ and $g_4$):
\begin{equation}\label{vacuum solutions}
\;\;
\Phi = \frac{\tr\Phi}{N} \! \1 \, ,
\qquad\quad
\Phi = \0 \, ,
\qquad\quad
\Phi^2 = \frac{g_2}{2g_4} \! \1 \,.
\end{equation}
The first two solutions correspond to ordered and the disordered vacua, respectively, both of which also appear in commutative models. The third solution, the so-called striped or matrix vacuum, is unique to the noncommutative setting and features eigenvalues of both signs. This phase breaks translational symmetry and gives rise to spatial modulation of magnetization \cite{Castorina:2007za, Mejia-Diaz:2014lza, Ambjorn:2002nj}. These eigenvalue configurations are visualized in Figure \ref{figure:01}, where they are also referred to as 1-cut symmetric (S1), 2-cut (M2\footnote{
In the notation M2, \enquote{M} denotes the matrix phase and \enquote{2} indicates a two-cut configuration of the eigenvalues.}, which can be purely symmetric i.e. S2),  and 1-cut asymmetric phases (A1). 

\begin{figure}[t]
\centering 
\includegraphics[width=0.49\textwidth]{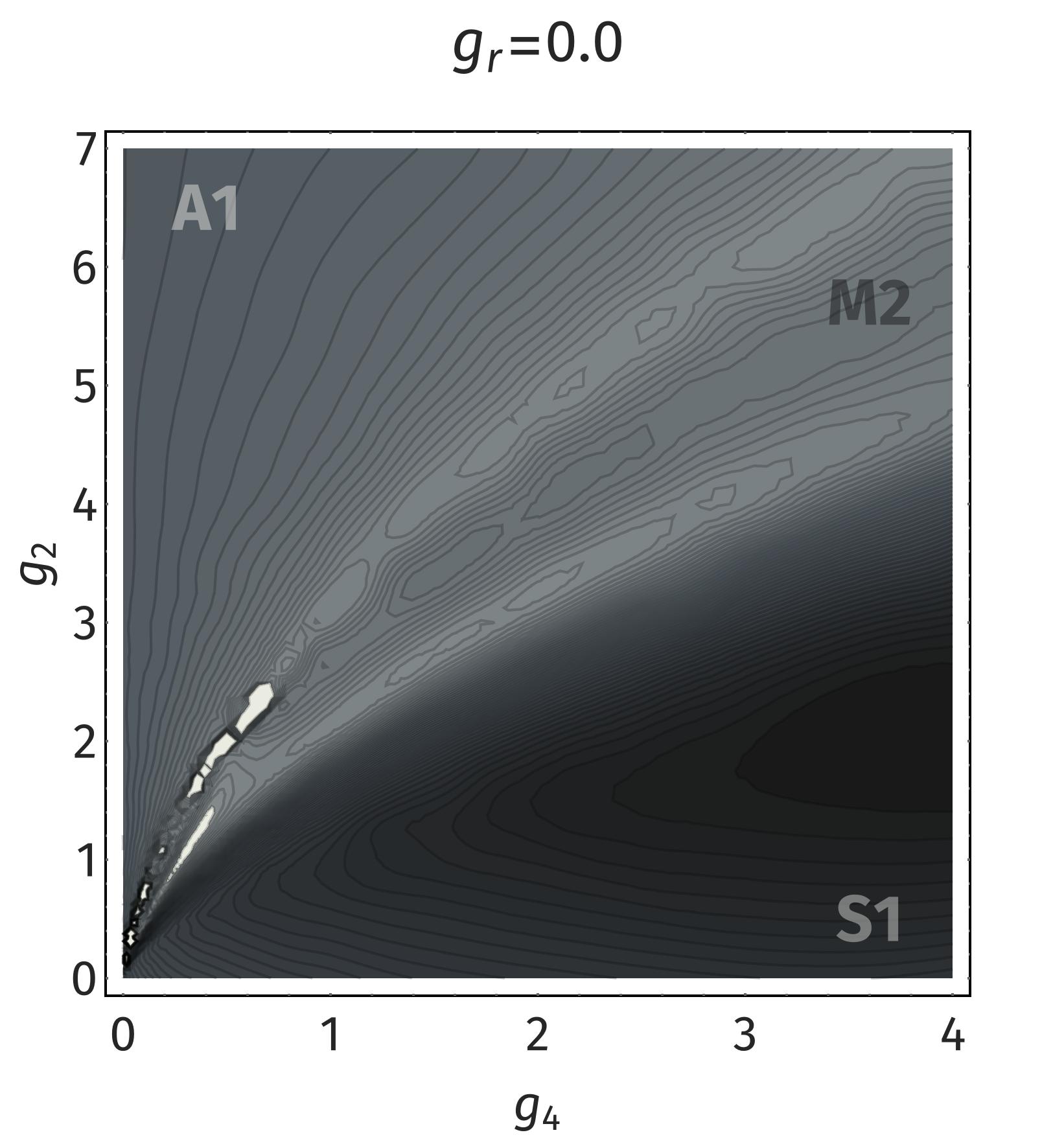}
\includegraphics[width=0.49\textwidth]{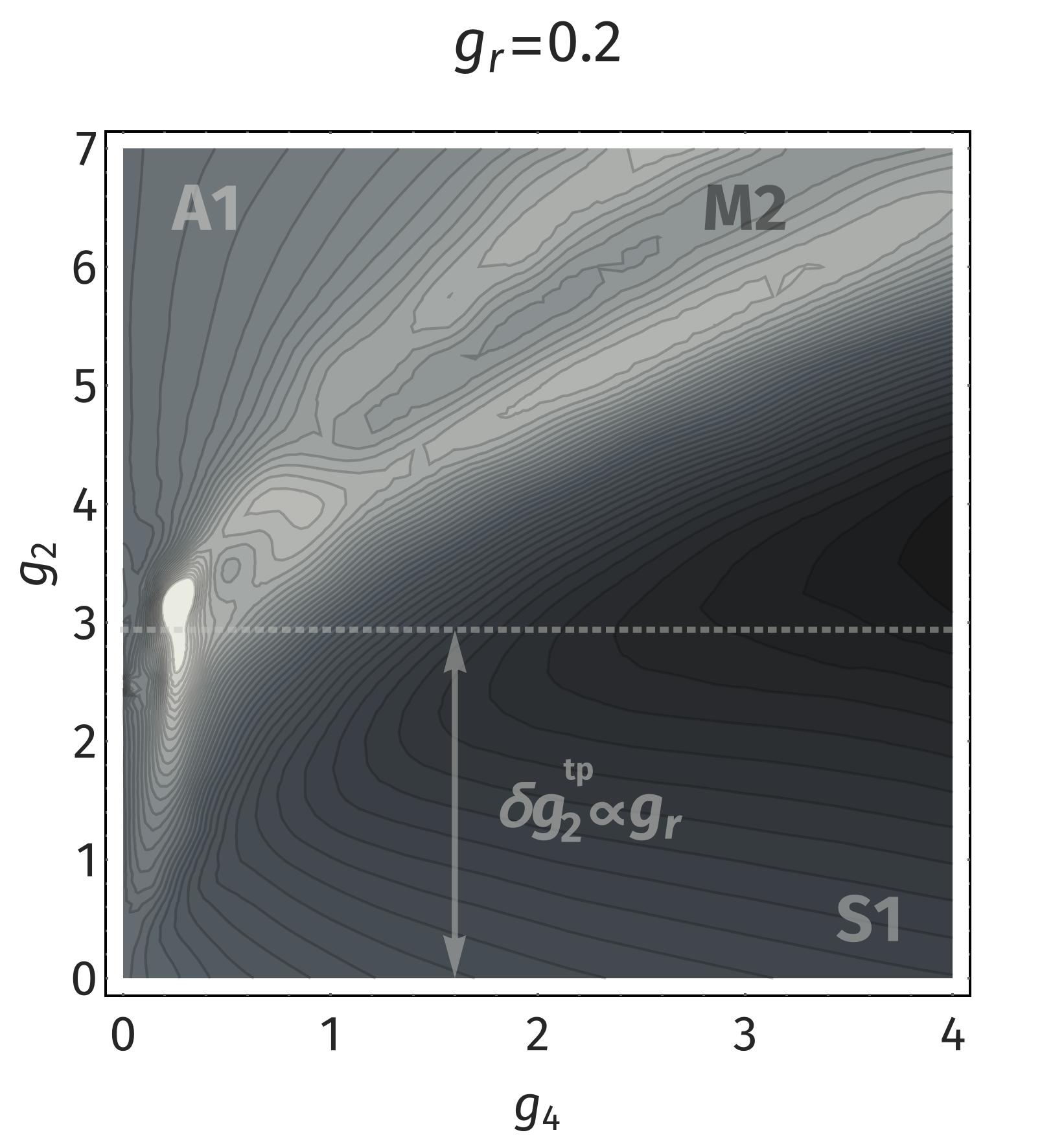}
\caption{
Phase diagrams of the $N=24$ matrix GW model without (left panel) and with
(right panel) the curvature term. Darker regions correspond to lower
values of the specific heat, $\Var S_\text{\tiny GW}^\text{\tiny M}/N^2$,
while lighter regions indicate higher values. The bright stripes mark
the phase transition lines, with phase labels (abbreviations) indicated
in the plot. Note that the triple point in the right-hand panel is
shifted by $\delta g_2^\text{tp} \approx 16 g_r = 16 \times 0.2 = 3.2$
relative to the left-hand panel.
}
\label{figure:02}
\end{figure}

Figure \ref{figure:02} presents\footnote{Figure \ref{figure:02} is adapted from our previous figures in \cite{Prekrat:2021uos} and \cite{Prekrat:2022sir}.} the overall structure of the phase diagram, obtained using HMC simulations \cite{Ydri:2015zba,betancourt2018conceptual}. This structure is similar to that found in other fuzzy spaces, such as the fuzzy sphere \cite{Kovacik:2018thy}. Notably, inclusion of the curvature term shifts the phase boundaries toward larger values of the mass parameter $g_2$. This is particularly visible in the displacement of the triple point, $\delta g_2^\text{tp}$, which we found numerically \cite{Prekrat:2021uos} to scale proportionally with the curvature coupling $g_r$. The resulting suppression of the striped phase has important consequences for the renormalization of the GW model.

To demonstrate this, let us look at the mass renormalization in the GW model given by \cite{Vinas:2014exa}
\begin{equation}
    \delta m^2_\text{ren} = \frac{\lambda}{12\pi(1+\Omega^2)}\ln\frac{\Lambda^2\theta}{\Omega} \, ,
\end{equation}
where the matrix size $N$ acts as the UV cutoff via ${\Lambda^2 \propto N}$ \cite{Grosse:2003nw}. Using the qualitative correspondences ${m \leftrightarrow G_2}$ and ${\Omega \leftrightarrow g_r}$  \cite{Prekrat:2021uos}, the renormalization shift in the matrix model becomes \cite{Prekrat:2021uos}
\begin{equation}
    \delta G_2^\text{ren} \sim -\ln N \, .
\end{equation}
This implies that the bare coupling $G_2$ must shift by 
\begin{equation}
    \abs{\delta G_2^\text{\,ren}}  \sim \ln N 
\end{equation}
to absorb quantum corrections. However, this is smaller than the shift of the triple point:
\begin{equation}
    \delta G_2^\text{tp} 
    = N \delta g_2^\text{tp}
    \sim N g_r \, ,
\end{equation}
which is the lowest-most-$g_2$ point of the striped phase (Figure \ref{figure:02}). As a result, the bare mass parameter must lie outside the striped phase (which is responsible for UV/IR mixing), and instead within the disordered phase with a trivial vacuum. The same applies to the renormalized version of the $\lambda\phi^4_\star$ model (denoted by $\lambda\phi^4_\text{\tiny GW}$), which is obtained \cite{Grosse:2003nw,Wulkenhaar:habilitation2004} as the limit of a sequence of GW models where the curvature coupling vanishes like
\begin{equation}
    \Omega \sim \frac{1}{\ln N} \to 0 \, .
\end{equation}
Further discussion on this topic can be found in \cite{Prekrat:2023fts}. In contrast, this protective shift does not occur in the original nonrenormalizable $\lambda\phi^4_\star$ model without the curvature term, whose triple point remains fixed at the origin in the large $N$ limit \cite{Prekrat:2021uos}, as seen in the ${g_r=0}$ plot in Figure \ref{figure:02}. 

It is also worth noting that changes to the kinetic term can induce similar shifts in the triple point position \cite{Subjakova:2020shi} and potentially resolve UV/IR mixing issues \cite{Dolan:2001gn}.

\section{Effective Action \& Transition Lines}

Since the link between the GW model’s renormalizability and the suppression of the striped phase hinges on a numerically observed shift in the phase diagram, it is important to confirm this shift analytically. Initial steps toward this goal were made in \cite{Prekrat:2022sir}, where the effective action up to $O(g_r^4)$ and the transition line between the disordered and striped phases were derived for the GW model without the kinetic term. 

As shown in \cite{Prekrat:2022sir}, the approximate analytical transition lines agree well with numerical data in the strong coupling regime. However, a more precise determination of the starting point (the triple point) requires higher-order approximations. In particular, we need to compare different perturbative orders of the effective action to verify whether the turning points of these transition lines converge in the perturbative regime. For this purpose, we derive the $O(g_r^6)$ effective action in this section.

To outline the approach, we recall from Figure \ref{figure:01} that the phase structure depends solely on the distribution of eigenvalues of the matrix field. Hence, we must integrate out the non-eigenvalue (angular) degrees of freedom to obtain the effective action. Starting with the decomposition ${\Phi = U \Lambda U^\dag}$ where $\Lambda$ is diagonal matrix of field eigenvalues and $U$ is unitary matrix, the action becomes:
\begin{equation}
    S_\text{\tiny GW}^\text{\tiny M} = N \tr\!\qty(
    (U \Lambda U^\dag) \mathcal{K} (U\Lambda U^\dag)
    - g_rRU\Lambda^2U^\dag
    - g_2\Lambda^2 + g_4\Lambda^4
    ) \, .
\end{equation}
However, integration over the unitary group is analytically intractable in full generality. Since, as demonstrated in Figure \ref{figure:02}, the shift of the triple point is entirely driven by the curvature term, we simplify the analysis by neglecting the kinetic term for the remainder of this paper and work with
\begin{equation}
    S = N \tr\!\qty(
    - g_rRU\Lambda^2U^\dag
    - g_2\Lambda^2 + g_4\Lambda^4
    ) \, ,
\end{equation}
that is, with the action
\begin{equation}\label{GW-without-K}
    S = N\tr\!\lr{
    - g_r R\Phi^2
    - g_2\Phi^2 + g_4\Phi^4
    } \, .
\end{equation}
This approximation is also supported by additional numerical evidence: for small $g_4$, which are precisely those relevant for the triple point, the eigenvalue distribution closely follows the configuration \cite{Prekrat:2023thesis} 
\begin{equation}\label{phiR}
    \Phi^2_R = \frac{g_2\!\1\!+g_rR}{2g_4} \, ,
\end{equation} 
which solves the EOM \eqref{EOM} when the kinetic term is omitted. The domain of existence of this solution imposes a lower bound on the triple point:
\begin{equation}
    g_2^\text{tp} \geq 16g_r \, .
\end{equation}
Conversely, replacing $R$ by its extremal eigenvalues provides an upper bound \cite{Prekrat:2020ptq}:
\begin{equation}
    g_2^\text{tp} \leq 16g_r \, .
\end{equation}
Taken together, they fix the triple point exactly at:
\begin{equation}
    g_2^\text{tp} = 16g_r \, . 
\end{equation}
Numerical simulations confirm that both bounds are indeed saturated.

In this simplified setting, we proceed to compute the effective action
from the partition function:
\begin{equation}
    Z = \int[\mathrm{d}\Phi] \, e^{-S} = 
    \int [\mathrm{d}\Lambda] \, \Delta^{\!2}\!(\Lambda) \, 
    e^{-N\tr\left(- g_2\Lambda^2 + g_4\Lambda^4\right)}
    \int [\mathrm{d}U]\,e^{\, g_r N\tr\lr{URU^\dagger \Lambda^2}}
    \, ,\label{partition}
\end{equation} 
where $\Delta(\Lambda)$ is the Vandermonde determinant:
\begin{equation}
    \Delta(\Lambda) = \prod_{1 \leq i < j \leq N} \!\!\! (\lambda_j - \lambda_i) \, ,
    \qquad
    \Lambda = \diag \lambda_i \, .\label{vandermonde}
\end{equation}
Following the method of \cite{Kanomata:2022pdo,Kanomata:2023mni}, the 
unitary group integral evaluates to \cite{Prekrat:2022sir}
\begin{equation}
Z= \int [\mathrm{d}\Lambda] \, 
    \frac{\Delta^2(\Lambda)}{\Delta(\Lambda^2)} \,
    e^{-N\tr(-g_2\Lambda^2 - g_r R\Lambda^2 + g_4 \Lambda^4)} \, ,  
\end{equation}
but since $\Delta(\Lambda^2)$ is not sign-definite, it cannot be absorbed into the effective action.
Because this direct route does not lead to the desired result, we instead turn to a perturbative approach, starting from the general integral
\begin{equation}
     I = \int\limits_{{\rm U}(N)} \!\! [\mathrm{d}U]\, e^{t\tr (UAU^\dag B)}
\end{equation}
for Hermitian matrices $A$ and $B$, with a normalized Haar measure.

This integral defines corrections $\delta S$ to the effective action:
\begin{equation}
    S_\text{eff} = -g_2 N \tr\!\Lambda^2 + g_4 N \tr\!\Lambda^4 
                 -\ln\Delta^2(\Lambda) + \delta S \, ,
\end{equation}
where
\begin{equation}
    I = \exp\qty(-\delta S)
    = \exp\qty(-\sum_{n=1}^\infty \frac{t^n}{n!}S_n) \, .
\end{equation}
Using the expansion:
\begin{equation}
     I = 1 + \sum_{n=1}^\infty \frac{t^n}{n!}\,I_n \, ,
     \qquad
     I_n = \int [\mathrm{d}U] \tr^n\! \qty(UAU^\dag B) \, ,
     \label{t:expansion}
\end{equation}
we obtain first few recursive formulas for $S_n$:

\begin{align}
    S_1^{} & = - I_1^{} \, ,
    \\
    S_2^{} & = S_1^2 - I_2^{} \, , 
    \\    
    S_3^{} & = - S_1^3 + 3 S_1^{} S_2^{} - I_3^{} \, ,
    \\
    S_4^{} & = S_1^4 - 6 S_1^2 S_2^{} + 3 S_2^2 + 4 S_1^{} S_3^{} - I_4^{} \, ,
    \\
    S_5^{} & = -S_1^5 +10 S_1^3 S_2^{} - 10 S_1^2 S_3^{} - 15 S_1^{} S_2^2 + 5 S_1^{} S_4^{} + 10 S_2^{} S_3^{} - I_5^{} \, ,
\end{align}
and
\begin{multline}
    \, S_6^{} = S_1^6 - 15 S_1^4 S_2^{} + 45 S_1^2 S_2^2 - 15 S_2^3 + 20 S_1^3 S_3^{} - 
               60 S_1^{} S_2^{} S_3^{} \, +
            \\
            + 10 S_3^2 - 15 S_1^2 S_4^{} + 15 S_2^{} S_4^{} + 
               6 S_1^{} S_5^{} - I_6^{} \, . \,
\end{multline}

In \cite{Prekrat:2024rmq}, we found $S_n$ and $I_n$ up to the 6th order with the help of the RTNI\footnote{In the meantime, an updated version of RTNI has been released \cite{Fukuda:2023xmz}.} computing package \cite{Fukuda:2019pzs}. We also proved that all odd orders of $S_n$ disappear, due to equidistant eigenvalues of the curvature and its symmetry w.r.t. the anti-diagonal. The resulting effective action up to $O(g_r^6)$ is:

\begin{align}\label{Seff}
    S_\text{eff} = 
    &
    - \qty(g_2 - 8g_r)N\tr\!\Lambda^2 
    + \qty(g_4 - \frac{32}{3}g_r^2)N\tr\!\Lambda^4 
    + \frac{32}{3}g_r^2\tr^2\!\!\Lambda^2
    \\
    &
    + \frac{1024}{45}g_r^4 N\tr\!\Lambda^8  
    + \frac{1024}{15}g_r^4\tr^2\!\!\Lambda^4
    - \frac{4096}{45}g_r^4\tr\!\Lambda^6\tr\!\Lambda^2
    \notag
    \hspace{18pt}
    \\
    &
    - \frac{262144}{2835} g_r^6 N\tr\!\Lambda^{12}
    + \frac{524288}{945} g_r^6 \tr\!\Lambda^2 \tr\!\Lambda^{10}
    - \frac{262144}{189} g_r^6 \tr\!\Lambda^4 \tr\!\Lambda^8
    + \frac{524288}{567} g_r^6 \tr^2\!\!\Lambda^6 
    \notag
    \\
    &- \ln\Delta^2(\Lambda) \, .
    \notag 
\end{align}   
Alternatively, writing this in powers of mass-shift parameter $8g_r$ (normalized trace of the curvature), the expansion becomes even more transparent:
\begin{align}\label{Seff-v2}
    S_\text{eff} = 
    &
    - \qty(g_2 - 8g_r)N\tr\!\Lambda^2 
    + \qty(g_4 - \frac{1}{6}(8g_r)^2)N\tr\!\Lambda^4 
    + \frac{1}{6}(8g_r)^2\tr^2\!\!\Lambda^2
    \\
    &
    + \frac{1}{180}(8g_r)^4 N\tr\!\Lambda^8  
    + \frac{1}{60}(8g_r)^4\tr^2\!\!\Lambda^4
    - \frac{1}{45}(8g_r)^4\tr\!\Lambda^6\tr\!\Lambda^2
    \notag
    \hspace{18pt}
    \\
    &
    - \frac{1}{2835} (8g_r)^6 N\tr\!\Lambda^{12}
    + \frac{2}{945} (8g_r)^6 \tr\!\Lambda^2 \tr\!\Lambda^{10}
    - \frac{1}{189} (8g_r)^6 \tr\!\Lambda^4 \tr\!\Lambda^8
    + \frac{2}{567} (8g_r)^6 \tr^2\!\!\Lambda^6 
    \notag
    \\
    &- \ln\Delta^2(\Lambda) \, .
    \notag
\end{align}   

\begin{figure}[t]
    \centering
    \includegraphics[width=0.90\textwidth]{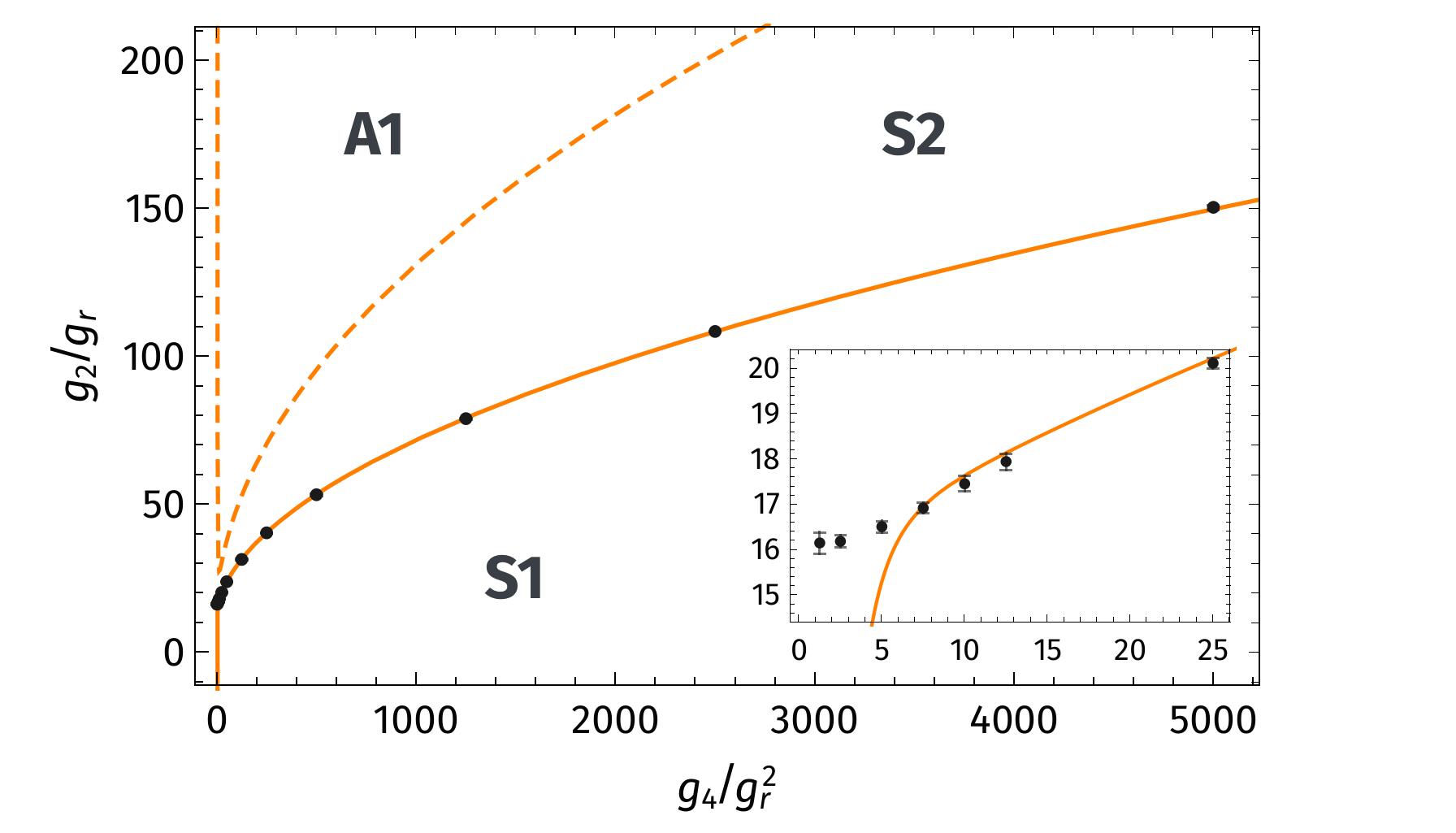}
    \caption{
    Phase transition lines obtained from the sixth-order effective action. Black dots indicate HMC simulation results for the action $S$ at ${g_r = 0.02}$, extrapolated to the ${N \to \infty}$ limit. The dashed line denotes the S2/A1 transition, which is absent in this model.}
    \label{figure:03}
\end{figure}

\begin{figure}[t]
    \centering
    \hspace{5pt}
    \includegraphics[width=0.96\textwidth]{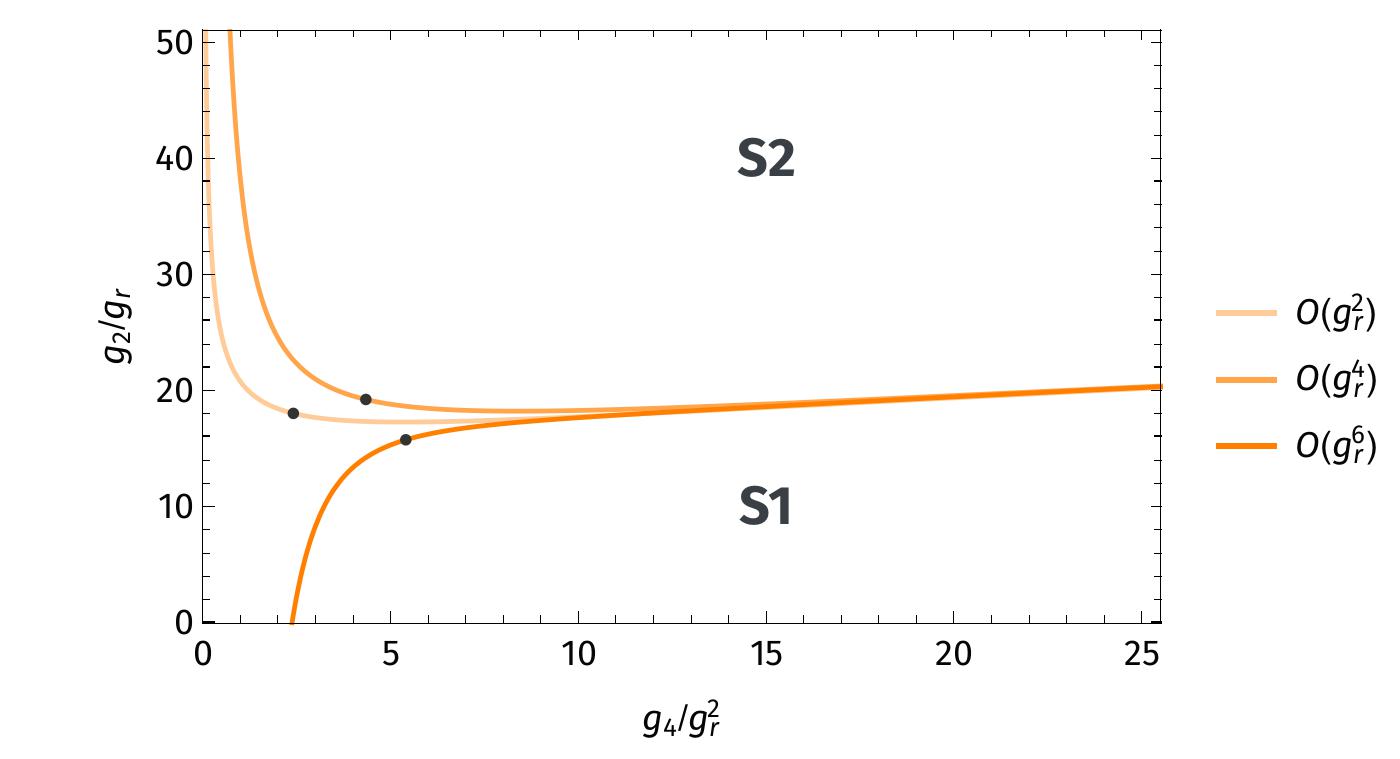}
    \caption{
    Comparison of the turning points predicted by the second-, fourth-, and sixth-order approximations to the exact S1/S2 transition line \eqref{S1-S2-line} of the model.}
    \label{figure:04}
\end{figure}

This effective action can now be used to derive\footnote{
An informative overview of the derivation of the eigenvalue distribution and the possible classes of solutions can be found in \cite{Subjakova:2020prh}.} the eigenvalue distribution and determine the corrections to the phase transition lines, leading to the better analytical estimation of the shift of the triple point. We employ three complementary strategies to extract physical predictions:
\begin{enumerate}
    \item Analytical expansion of the ${N \to \infty}$ transition lines in powers of $g_r$, based on the derived effective action $S_\text{eff}$.
    \item Numerical analysis of the continuum limit of multitrace matrix models obtained by truncating $S_\text{eff}$ at several finite orders in $g_r$.   
    \item HMC simulations of the full effects of the action $S$, given by \eqref{GW-without-K}, to locate the transition line for finite $N$, followed by extrapolation to ${N \to \infty}$.
\end{enumerate}
Skipping ahead to the results, we list the analytical transition lines up to $O(g_r^6)$:
\begin{itemize}
    \item Transition from S1 to S2 phase:
\begin{equation}\label{S1-S2-line}
    g_2 = 2\sqrt{g_4} 
    + 8g_r 
    + \frac{32}{3}\frac{g_r^2}{\sqrt{g_4}} 
    + \frac{256}{15}\frac{g_r^4}{g_4\sqrt{g_4}} 
    - \frac{4096}{21}\frac{g_r^6}{g_4^2\sqrt{g_4}}\, .
\end{equation}
    \item(Unrealized) transition from S2 to A1 phase:
\begin{equation}\label{S2-A1-line}
    g_2 = \sqrt{15} \sqrt{g_4} 
    + 8g_r 
    + \frac{592}{9\sqrt{15}}\frac{g_r^2}{\sqrt{g_4}} 
    - \frac{503168}{1125\sqrt{15}}\frac{g_r^4}{g_4\sqrt{g_4}} 
    + \frac{1581033472}{212625\sqrt{15}}\frac{g_r^6}{g_4^2\sqrt{g_4}}\, .
\end{equation}
\end{itemize}
To simplify, we define dimensionless deformation parameters (relative to the ${g_r = 0}$ transition lines): 
\begin{equation}
    \epsilon = \frac{8 g_r}{2\sqrt{g_4}} \ ,
    \qquad
    \bar{\epsilon} = \frac{8 g_r}{\sqrt{15g_4}} \ ,
\end{equation}
so that the above become:
\begin{itemize}
    \item Transition from S1 to S2 phase:
\begin{equation}\label{S1-S2-line-v2}
    \frac{g_2}{2\sqrt{g_4}} = 1 
    + \epsilon 
    + \frac{1}{3}\epsilon^2
    + \frac{1}{30}\epsilon^4 
    - \frac{1}{42}\epsilon^6 \, .
\end{equation}
    \item(Unrealized) transition from S2 to A1 phase:
\begin{equation}\label{S2-A1-line-v2}
    \frac{g_2}{\sqrt{15 g_4}} = 1 
    + \bar{\epsilon} 
    + \frac{37}{36}\bar{\epsilon}^2
    - \frac{3931}{2400}\bar{\epsilon}^4 
    + \frac{771989}{120960}\bar{\epsilon}^6 \, .
\end{equation}
\end{itemize}
The analytical transition lines, alongside the HMC results, are presented in Figure \ref{figure:03}. However, these expressions alone do not suffice to accurately pinpoint the location of the triple point. This limitation is evident in Figure \ref{figure:04}, where the turning points of the second-, fourth-, and sixth-order approximations to the S1/S2 transition line \eqref{S1-S2-line} exhibit erratic behavior, \enquote{jumping around} rather than converging monotonically to a single point. Moreover, these points deviate from the anticipated triple point position at ${(g_4, g_2) = (0,16g_r)}$. Consequently, any convergence can only be inferred by analyzing higher-order approximations. The same holds for the turning points of the would-be S2/A1 transition line.

We conclude the section by noting that analogous expansion techniques have been initiated for the kinetic term's contribution to the effective action \cite{Benedek:2023stconf}. The first nontrivial terms of this expansion are:
\begin{multline}
    S_\text{eff}^\text{kin}(\Lambda) = 
      N\tr\!\Lambda^2 - \tr^2\!\!\Lambda
    + \frac{97}{120} N \tr\!\Lambda^4
    - \frac{565}{120N^2} \tr^4\!\!\Lambda
    \, +
    \\
    + \frac{113}{12N} \tr^2\!\!\Lambda \tr\!\Lambda^2
    - \frac{137}{60} \tr^2\!\!\Lambda^2
    - \frac{97}{30} \tr\!\Lambda \tr\!\Lambda^3
    \, ,
\end{multline}
and will be explored in future work.

We will now provide a detailed explanation of the three aforementioned strategies employed to obtain our results.

\section{Review of the Matrix Model Description}\label{sec5}

\subsection{Eigenvalue Distribution}

The phase transitions observed in our model are associated with topological changes in the support of the eigenvalue distribution. To study these transitions, we first derive the eigenvalue distribution for matrix models.

We consider  ${N\times N}$ Hermitian matrices with the Dyson integration measure $[\mathrm{d}\Phi]$, and define the probability measure as $e^{-N^2 \mathcal{S}(\Phi)}$. In this section, for convenience, we work with a rescaled action $\mathcal{S}$, specifically, ${\mathcal{S} = O(1)}$ while previously ${S = O(N^2)}$. We also consider a more general form of the potential $V(\Phi)$. The expectation value of a function $f(\Phi)$ is then computed from the partition function~\eqref{partition} as
\begin{align}
\avg{f} \,=\,\frac{1}{Z}\int [\mathrm{d}\Phi] e^{-N^2 \mathcal{S}(\Phi)} f(\Phi)\ ,
\qquad 
Z \,=\,\int [\mathrm{d}\Phi] e^{-N^2 \mathcal{S}(\Phi)}\ ,\label{f}
\end{align}
with the variance given by
\begin{equation}
    \Var f = \avg{f^2} - \smash{\avg{f^{\phantom{2}\!\!\!}}}^2.
\end{equation}
The most general single-trace action can be written as:
\begin{align}
\mathcal{S}(\Phi)\,=\, \frac{1}{N}\tr V(\Phi)\, ,
\qquad 
V(\Phi)\,=\,\sum_{n=0}^{N}g_n\,\Phi^n,
\qquad 
g_n\in\mathds{R}\ .\label{S}
\end{align}
Here, the $1/N$ factor ensures that $\mathcal{S}(\Phi)$ remains of order one when the trace\footnote{The constant term in the definition of $V(\Phi)$ can be left out because it will cancel thanks to \eqref{f}, anyway. We can get rid of the linear term as well by means of shifting the matrices with a constant matrix.} is taken. Upon diagonalizing the Hermitian matrix $\Phi$, we write ${\Phi=U\Lambda U^{\dag}}$, where $\Lambda = \diag \lambda_i$ contains eigenvalues of $\Phi$ and ${U\in\text{U}(N)}$. 
The Jacobian of this transformation introduces the aforementioned Vandermonde determinant \eqref{vandermonde}:
\begin{align}
[\mathrm{d}\Phi]=[\mathrm{d}U]\lr{ \prod_{i=1}^N \mathrm{d}\lambda_i}\lr{ \prod_{i<j}\lr{\lambda_i-\lambda_j}^2} \, .
\end{align}
Assuming ${f(\Phi)}$ is invariant under unitary conjugation (i.e. depends only on eigenvalues), the integral with respect to the Haar measure $[\mathrm{d}U]$ becomes trivial, and the expectation value reduces to: 
\begin{align}
\avg{f}
\,=\,
\frac{1}{Z} \int \lr{ \prod_{i=1}^N \mathrm{d}\lambda_i} 
\exp\lr{-N^2 \slr{\mathcal{S}(\Lambda)-\frac{2}{N^2}\sum\limits_{i<j}\text{ln}\lvert\lambda_i -\lambda_j\rvert}} f(\lambda_i )\, .\label{fn}
\end{align}
We define the quantity in square brackets as the free energy \cite{Tekel:2015uza}:
\begin{align}
\mathcal{F}\,=\, \mathcal{S}(\Lambda)-\frac{2}{N^2}\sum_{i<j}\text{ln}\lvert\lambda_i -\lambda_j\rvert\, .\label{FREE}
\end{align}
From this point onward, we focus exclusively on the continuum limit ${N\rightarrow\infty}$, where the theory becomes more amenable to analytical treatment.
In the large-$N$ limit, only the configuration of eigenvalues $\widetilde\lambda_i$ that globally minimizes $\mathcal{F}$ significantly contributes. This leads to the saddle point equation for the singletrace action \eqref{S}: 
\begin{align}
V'(\widetilde\lambda_i)-\frac{2}{N}\sum_{j\neq i} \frac{1}{\widetilde\lambda_i-\widetilde\lambda_j}\,=\,0\, .\label{dspa}
\end{align}
This equation describes the equilibrium of repulsive eigenvalue \enquote{particles} in an external potential $V(\Phi)$. To analyze the system further, we define:
\begin{itemize}
    \item The eigenvalue density:
    \begin{equation}\label{def:distribution}
    \rho(\lambda) = \frac{1}{N}\sum_{i=1}^N\delta\lr{\lambda-\widetilde\lambda_i} \, .  
    \end{equation}
    \item The resolvent:
    \begin{equation}\label{def:resolvent}
    \omega(z) = \frac{1}{N}\sum_{i=1}^N\frac{1}{z-\widetilde\lambda_i} \, .
    \end{equation}
    \item The distribution moments:
    \begin{equation}\label{def:moments}
    c_n = \frac{1}{N}\tr\Phi^n \, .
    \end{equation}
\end{itemize}
In the $N \to \infty$ limit, the stable configuration $\widetilde\lambda_i$ becomes continuous distribution function $\rho(\lambda)$ and sums become integrals:
\begin{align}
\lim_{N\rightarrow\infty}\frac{1}{N}\sum_{i=1}^N f(\tilde \lambda_i)\longrightarrow \int_\C \mathrm{d}\lambda'\, \rho(\lambda')\, f(\lambda')\, ,
\label{trans}
\end{align}
where $\C$ is the support of the distribution. The resolvent and moments become:
\begin{align}
\omega(z) \,=\, \int_\C \mathrm{d}\lambda'\, \frac{\rho(\lambda')}{z-\lambda'} \, ,
\qquad\qquad 
c_n = \int_\C \mathrm{d}\lambda'\,\rho(\lambda')\lambda'^n \, .
\label{oc}
\end{align}
The saddle point equation \eqref{dspa}  transforms into the singular integral equation:
\begin{align}
V'(\lambda)- 2 \cdot \pv\!\int_\C \mathrm{d}\lambda'\, \frac{\rho(\lambda')}{\lambda-\lambda'}\,=\,0\, ,
\label{vspe}
\end{align}
where $\pv$ denotes the principal value. Using the Sokhotski–Plemelj formula \cite{Tekel:2015uza},
\begin{align}
\pv\int_\C \mathrm{d}\lambda'\, \frac{\rho(\lambda')}{\lambda-\lambda'}\,=\,\lim_{\varepsilon\rightarrow0^+}\slr{\omega(\lambda\pm\text{i}\varepsilon)\pm\text{i}\pi\rho(\lambda)}  \, ,
\label{}
\end{align}
we express \eqref{vspe} and  $\rho(\lambda)$ as:
\begin{align}
V'(\lambda)\,=\,\lim_{\varepsilon\rightarrow0^+}\slr{\omega(\lambda+\text{i}\varepsilon)+\omega(\lambda-\text{i}\varepsilon)} \, ,
\label{}
\end{align}
\begin{align}
\rho(\lambda)\,=\,\frac{1}{2\pi\text{i}}\,\lim_{\varepsilon\rightarrow0^+}\slr{\omega(\lambda+\text{i}\varepsilon)-\omega(\lambda-\text{i}\varepsilon)} \, .
\label{rholambda}
\end{align}
Squaring the resolvent, neglecting all the subdominant terms and employing the saddle point equation, one finds a quadratic equation with general solution:
\begin{align}
\omega(z)=\frac12\slr{V'(z)-\lvert H(z)\rvert\sqrt{\sigma(z)}} \, .
\label{omega2} 
\end{align}
Here, $\sigma(z)$ is an even polynomial whose roots determine the support $\C$, while $\lvert H(z)\rvert$ does not contribute to the discontinuity.

We focus on three typical cases of eigenvalue support:
\begin{itemize}
  \item Symmetric 1-cut (S1):
  \begin{equation}
  \sigma_{\text{S1}}(\lambda) = \prod_{\pm}\lr{\lambda \pm \sqrt{\delta}}\, ,
  \qquad
  \C_{\text{S1}} = \lr{-\sqrt{\delta}\, ,\,\sqrt{\delta}\,} \, .
  \end{equation}
  \item Symmetric\footnote{We have also considered asymmetric 2-cut solutions of our model but could not find any.} 2-cut (S2):
  \\
  \begin{multline}
  \sigma_{\text{S2}}(\lambda) = 
  \prod_{\pm, \pm} \lr{\lambda \pm \sqrt{D \pm \delta}}\, ,
  \\
  \C_{\text{S2}} = 
  \lr{\,-\sqrt{D+\delta} \, ,\, -\sqrt{D-\delta}\,}
  \,\cup\,
  \lr{\,\sqrt{D-\delta} \, ,\, \sqrt{D+\delta}\,}\, . 
  \label{Comega2}
  \end{multline} 
  \item Asymmetric 1-cut (A1):
  \begin{equation}
  \sigma_{\text{A1}}(\lambda) =
  \prod_{\pm} \lr{\lambda - \lr{D \pm \sqrt{\delta}}\,} \, ,
  \qquad
  \C_{\text{A1}} = \lr{ D-\sqrt{\delta} \, , \, D+\sqrt{\delta}\,} \, .
  \end{equation}
\end{itemize}
Under condition
\begin{align}
\omega(z)\sim\frac{1}{z}\, ,
\qquad
\lvert z\rvert\rightarrow\infty\ 
\label{omega3}
\end{align}
the polynomial part of \eqref{omega2} vanishes as ${\lvert z\rvert\rightarrow\infty}$, which determines both $\lvert H(z) \rvert$ and the endpoints of the support.
From \eqref{rholambda}, the eigenvalue density is then given by:
\begin{align}
\rho(\lambda)=\frac{1}{2\text{i}\pi}\lvert H(\lambda)\rvert\sqrt{\sigma(\lambda)}\, .
\label{} 
\end{align}
The moments $c_n$ can be extracted from the expansion of the resolvent using \eqref{oc}: 
\begin{align}
\omega(z)\,=\,\frac{1}{z}\int_\C \mathrm{d}\lambda'\, \sum_{n=0}^{\infty}\lr{\frac{\lambda'}{z}}^n\rho(\lambda')\,=\,\sum_{n=0}^{\infty}\frac{c_n}{z^{n+1}}\, .
\label{omega4}
\end{align}

\color{black}

\subsection{Free Energy for Single-Trace Action}\label{}

In the continuum limit, the free energy \eqref{FREE} becomes 
\begin{equation}
\mathcal{F}\,=\, \int\limits_\C \mathrm{d}\lambda\, \rho(\lambda)\, V(\lambda)
\quad - \;
\iint\limits_{\C\times\C\setminus \{\lambda=\tau\}}
\mathrm{d}\lambda \, \mathrm{d}\tau 
\rho(\lambda) \rho(\tau)\ln\lvert\lambda -\tau\rvert\, .
\label{FREE1}
\end{equation}
The double integral is challenging to evaluate directly, but we can employ a useful technique introduced in \cite{Tekel:2015uza}, which applies to symmetric distributions and the A1 phase. 

Specifically, although the free energy $\mathcal{F}$ has already been minimized in \eqref{dspa}, we now minimize the action augmented by a Lagrange multiplier $\xi$:
\begin{equation}
\mathcal{S}_{V}(\rho)\,=\, \mathcal{F}+\xi\lr{1-\int_\C \mathrm{d}\lambda\, \rho(\lambda)} \, , \label{}
\end{equation}
where the variation is taken with respect to $\rho(\lambda)$, ensuring the proper normalization of the eigenvalue density. This procedure yields the same value of the multiplier,
\begin{equation}
\xi\,=\, V(\lambda)-2\int_{\C}\mathrm{d}\tau  \rho(\tau)\ln\lvert\lambda -\tau\rvert\, ,\label{xi}
\end{equation}
for all ${\tau\in\C}$. Substituting this result back into \eqref{FREE1}, we arrive at a more elegant expression for the free energy:
\begin{equation}
\mathcal{F}\,=\,\frac12\lr{\int_\C \mathrm{d}\lambda\, \rho(\lambda)\, V(\lambda) +\xi} \, .\label{F}
\end{equation}

\subsection{Multitrace Term of Second Degree}\label{sec2.3}

In order to explore the effects of multitrace terms containing products of moments, such as those found in \eqref{Seff}, we will add an additional term to the singletrace action \eqref{S}:
\begin{align}
\mathcal{S}(\Phi)\,=\,\frac{1}{N}\tr V(\Phi)+t\,c_nc_m \, ,
\qquad
t\in\mathds{R}\ .
\label{multiSm}
\end{align}
The free energy of this multitrace action is:
\begin{align}
\mathcal{F}\,=\,\frac{1}{N}\tr V(\Phi)+tc_nc_m -\frac{2}{N^2}\sum_{i<j}\text{ln}\lvert\lambda_i -\lambda_j\rvert\, ,\label{FREEm}
\end{align}
yielding the saddle point equation:
\begin{align}
V'(\widetilde\lambda_i)+t\lr{c_nm\widetilde\lambda_i^{m-1}+c_mn\widetilde\lambda_i^{n-1}}-\frac{2}{N}\sum_{j\neq i} \frac{1}{\widetilde\lambda_i-\widetilde\lambda_j}\,=\,0\, .\label{spm}
\end{align}
This equation matches that of a singletrace model with the effective potential:
\begin{align}
V_{\text{eff}}(\Phi)\,=\,V(\Phi)+tc_n\Phi^m+tc_m\Phi^n \, ,
\label{Veff}
\end{align}
which has effective free energy:
\begin{align}
\mathcal{F}_{\text{eff}}\,=\,\frac{1}{N}\tr V(\Phi)+2tc_nc_m -\frac{2}{N^2}\sum_{i<j}\text{ln}\lvert\lambda_i -\lambda_j\rvert\, .
\label{FREEe}
\end{align}
The original free energy \eqref{FREEm} can then be written as:
\begin{align}
\mathcal{F}\,=\,\mathcal{F}_{\text{eff}}-tc_nc_m\, .\label{Feff}
\end{align}
To compute  $\mathcal{F}_{\text{eff}}$, the effective model \eqref{Veff} could in principle be solved using the standard techniques from earlier \cite{Tekel:2015uza}. 
However, since the parameters of the effective model depend on the moments, which in turn depend on the eigenvalue distribution, one must solve a system of self-consistent equations for both the moments and the support of the distribution. Recall that the moment-equations come from the expansion of the resolvent \eqref{omega4}, while the support-equations fix the endpoints of the cut(s).

\section{Phase Diagrams for Multitrace Submodels}\label{sec6}

We now analyze the phase structure of the matrix model defined by the effective action \eqref{Seff}, whose effective potential, following the definitions of the previous subsection, reads:
\begin{align}
V_{\text{eff}}(\Phi)=& -\left(g_2-8 g_r-\frac{64}{3}g_r^2c_2 +\frac{4096}{45}g_r^4c_6-\frac{524288}{945}g_r^6 c_{10} \right)\Phi^2+\nonumber\\
&+\left(g_4-\frac{32}{3} g_r^2+\frac{2048 }{15}g_r^4c_4-\frac{262144}{189} g_r^6c_8 \right)\Phi^4+\nonumber\\
&+\left(-\frac{4096 }{45}g_r^4c_2+\frac{1048576}{567}g_r^6 c_6 \right)\Phi^6+\nonumber\\
&+\left(\frac{1024}{45}g_r^4-\frac{262144}{189}g_r^6c_4\right)\Phi^8+\frac{524288 }{ 945}g_r^6c_2\Phi^{10}-\frac{262144}{2835}g_r^6\Phi^{12}\, .\label{VeffGW}
\end{align}
Our goal is to determine the phase diagram of this model, identifying the regions in the $(g_4,\,g_2)$ parameter space where different eigenvalue distribution topologies minimize the free energy at fixed $g_r$.

For each parameter pair $(g_4,\,g_2)$, we solve the saddle point equations for the eigenvalue distribution and identify the configuration that minimizes the free energy:
\begin{multline}
    \mathcal{F}= \mathcal{F}_\text{eff} 
    - \frac{32}{3}g_r^2c_2^2 
    - \frac{1024}{15}g_r^4c_4^2
    + \frac{4096}{45}g_r^4c_6c_2-
    \\
    - \frac{524288}{945} g_r^6 c_2c_{10}
    + \frac{262144}{189} g_r^6 c_4c_8
    - \frac{524288}{567} g_r^6 c_6^2
    \, .
\end{multline} 
We remind the reader that $\mathcal{F}_\text{eff}$ is the free energy corresponding to the potential $V_{\text{eff}}(\Phi)$ from \eqref{VeffGW} while $\mathcal{F}$ is the true free energy associated with the action $S_\text{eff}$ in \eqref{Seff}.
Only solutions corresponding to global minima are retained; all other possible solutions are disregarded.

Due to the complexity of the effective potential, the saddle point equations are solved numerically. We emphasize that this does not refer to the HMC simulation of the matrix model. 

To find the eigenvalue distribution for given couplings, we solve for:
\begin{itemize}
    \item the ${\deg H - 2}$ coefficients $h_n$ of the polynomial part of the resolvent,
    \item two support parameters $D$ and $\delta$, defining the centers and the radii of the cut(s),
    \item  and the moments $c_n$, which enter the effective potential coefficients.
\end{itemize}
More specifically, the coefficients $h_n$ are determined from \eqref{omega3}, which requires that all coefficients of powers of $z$ higher than $z^0$ in \eqref{omega2} vanish. 
Treating the moments $c_n$ as free parameters at this stage, matching the power series expansions of \eqref{omega2} and \eqref{omega4} yields a system of equations that is linear in the $c_n$. This linearity is not generic but follows from the fact that the effective action \eqref{Seff} is quadratic in the moments, while the resulting equations remain nonlinear in the support parameters. However, the moments themselves depend on the $h_n$ through the integrals in \eqref{oc} and also appear explicitly in the equations, leading to a self-consistent determination of the $c_n$.
The only remaining degrees of freedom are the support parameters, which are then fixed numerically by solving \eqref{omega3}, imposing that the coefficients at $z^0$ and  $z^{-1}$ take the values 0 and 1, respectively. This procedure ultimately establishes whether the solution corresponds to a S1 or a S2 configuration.

\begin{figure}[t]
    \centering
    \includegraphics[width=0.8\textwidth]{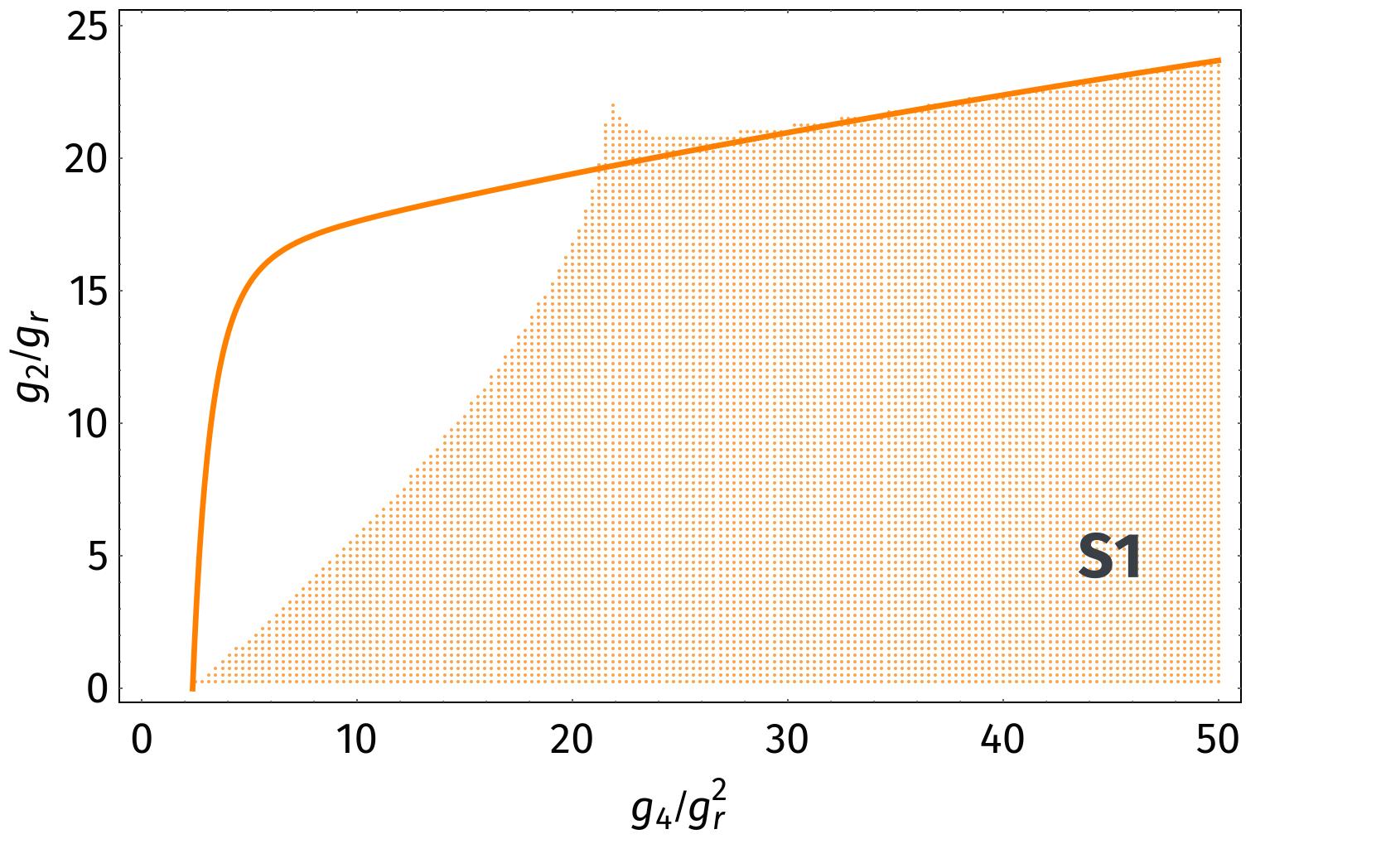}
    \caption{
    Densely distributed dots indicate the numerically confirmed S1-phase solutions of the $O(g_r^6)$ effective action \eqref{Seff} at ${g_r = 0.02}$. The solid line represents the corresponding $O(g_r^6)$ analytical S1/S2 transition line \eqref{S1-S2-line}.}
    \label{figure:05}
\end{figure}

We can also determine the support parameters analytically by expanding them in powers of $g_r$ and then solving the $z^0$ and $z^{-1}$ equations order by order, up to the highest power of $g_r$ appearing in the effective action. Once the eigenvalue distribution parameters have been obtained in this way, we turn to the relevant transition condition and solve it in terms of the action couplings. For example, if we wish to express the transition line $g_2$ as a function of $g_4$, we again expand this function in powers of $g_r$ and solve the condition order by order in $g_r$. In the case of the S1 to S2 transition, the appropriate condition is the splitting of the eigenvalue distribution support into two parts across the center, i.e.,
\begin{equation}
    \rho(g_i;\, \lambda = 0) = 0 \, ,
\end{equation}
while for S2 to A1 transition, the condition is that the distribution becomes negative at the inner edge of the support, meaning the polynomial part has a zero at that point, 
\begin{equation}
    H(D - \sqrt{\delta}) = 0 \, .
\end{equation}
This gives us the transition line equations \eqref{S1-S2-line-v2} and \eqref{S2-A1-line-v2}. 
The comparison between the numerical and analytical results mentioned in this section is illustrated in Figure \ref{figure:05} where they display an excellent agreement for larger values of $g_4$. However, the analytical expression starts to break down for smaller $g_4$ (in this case, near ${g_4 \approx 25 g_r^2}$), which is a general feature for all the studied orders of approximation.

\subsection{Phase Diagram for Second-Order Action}

We now apply the methods outlined in the previous section to investigate the simplest multitrace submodel---namely, the second-order effective action given by the first line of equation \eqref{Seff} (and equivalently \eqref{Seff-v2}), supplemented by the Vandermonde contribution. In this case, the self-consistent integral system simplifies enough to allow an exact solution once the phase transition conditions are imposed. We stress that, for the purposes of this section, we treat the second-order action not as an approximation but as a complete action in its own right. Consequently, the resulting expressions contain higher powers of $g_r$, and their second-order truncations represent second-order approximations of the exact transition lines.

Notably, in the spirit of \eqref{spm}, the eigenvalue distribution equation for the pure potential model with a $\tr^2 \Phi^2$ term is identical in form to that of the standard single-trace quartic model, but with effective couplings shifted by the new term. In particular, one finds modified mass and quartic couplings:
\begin{equation}\label{effpar}
   g_{2,\text{eff}} = g_2 - 8g_r - \frac{64}{3}g_r^2c_2\, ,
   \qquad \qquad
   g_{4,\text{eff}} = g_4 - \frac{32}{3}g_r^2\, .
\end{equation} 
Given the known exact solutions for the S1/S2 and S2/A1 transition lines in the pure potential model, we can use them to derive the corresponding exact transition lines in the presence of the $\tr^2 \Phi^2$ term.

\begin{figure}[t]
    \centering
    \includegraphics[width=0.99\textwidth]{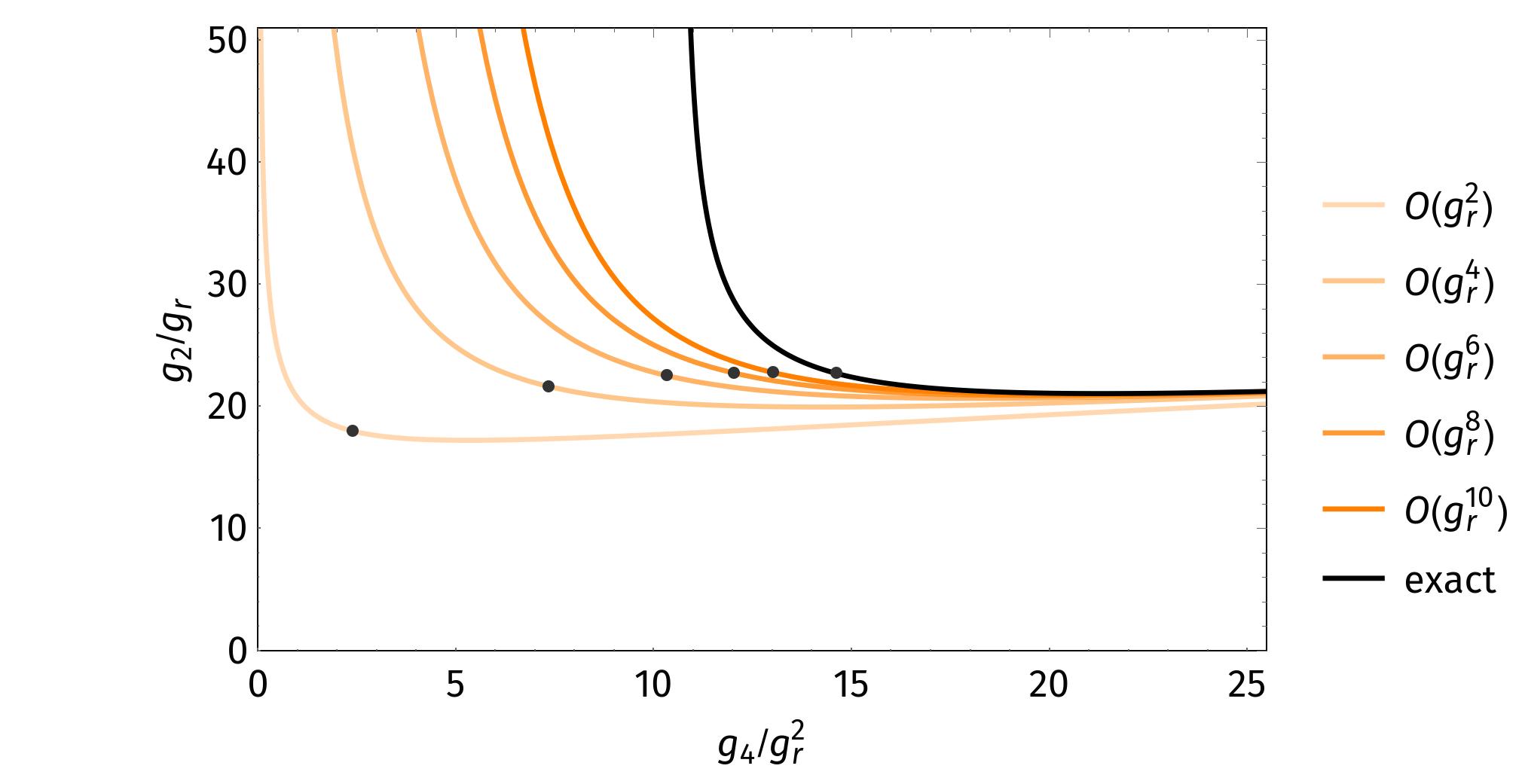}
    \caption{
    Analytical transition line of the second-order GW model without the kinetic term, shown for different orders of approximation in the curvature coupling $g_r$. The dots indicate the \enquote{turning points} of each curve, marking the locations of maximum arc curvature.}
    \label{figure:06}
\end{figure}

Focusing first on the S1 solution, the exact transition line between the S1 and S2 phases for the second-order effective action was previously obtained in \cite{Prekrat:2022sir}. It is given by:
\begin{align}
   g_2=2\sqrt{g_4-\frac{32}{3}g_r^2}+8g_r+\frac{\dfrac{64}{3}g_r^2}{\sqrt{g_4-\dfrac{32}{3}g_r^2}} \, ,\label{g22order}
\end{align} 
and shown in Figure \ref{figure:06}, which also displays successive perturbative approximations to this
transition line.
As mentioned in the Introduction, the turning points of these
approximations appear to converge monotonically toward the exact
result, which motivated the investigation of higher-order contributions to the effective action in the first place.


We now repeat the used method for the would-be S2/A1 transition line. For the general A1 solution, the support parameters (the $D$ center and the $\sqrt{\delta}$ radius of the support) are obtained by solving the $z^0$ and $z^{-1}$ conditions for the resolvent. This yields analytical expressions (as given in \cite{Tekel:2015uza}):
\begin{align}\label{Dd}
   D = \pm\frac{\sqrt{\displaystyle 3g_{2,\text{eff}}+2\sqrt{g_{2,\text{eff}}^2
     - 15g_{4,\text{eff}}}}}{\sqrt{10g_{4,\text{eff}}}} \, ,
   \qquad \quad
   \delta = 2\cdot\frac{g_{2,\text{eff}}
          - \sqrt{g_{2,\text{eff}}^2-15g_{4,\text{eff}}}}{15g_{4,\text{eff}}} \, .
\end{align} 
Requiring that the arguments of the square roots in \eqref{Dd} remain non-negative leads to the following transition line equation:
\begin{align}
   g_{2,\text{eff}}=\sqrt{15\, g_{4,\text{eff}}}\, .\label{transpure}
\end{align} 
Further on, the second moment $c_2$ of the A1 eigenvalue distribution satisfying \eqref{Dd} can be obtained directly from its definition:
\begin{equation}\label{c2A1}
c_2=\frac{\lr{g_{2,\text{eff}}-\sqrt{g_{2,\text{eff}}^2-15 g_{4,\text{eff}}}} 
\lr{41 g_{2,\text{eff}} \sqrt{g_{2,\text{eff}}^2-15 g_{4,\text{eff}}}+49 g_{2,\text{eff}}^2-120 g_{4,\text{eff}}}}{1350\, g_{4,\text{eff}}^2} \, .
\end{equation} 
Solving the combined equations \eqref{effpar}, \eqref{Dd}, \eqref{transpure} and \eqref{c2A1}, we obtain the exact would-be transition line between the S2 and A1 phases:
\begin{equation}\label{S2/A1-exact}
    g_2 = \sqrt{15}\sqrt{g_4-\frac{32}{3}g_r^2} + 8g_r + \frac{\dfrac{1312}{9} g_r^2}{\sqrt{15}\sqrt{g_4-\dfrac{32}{3}g_r^2}}\, .
\end{equation}

\begin{figure}[t]
    \centering
    \includegraphics[width=0.90\textwidth]{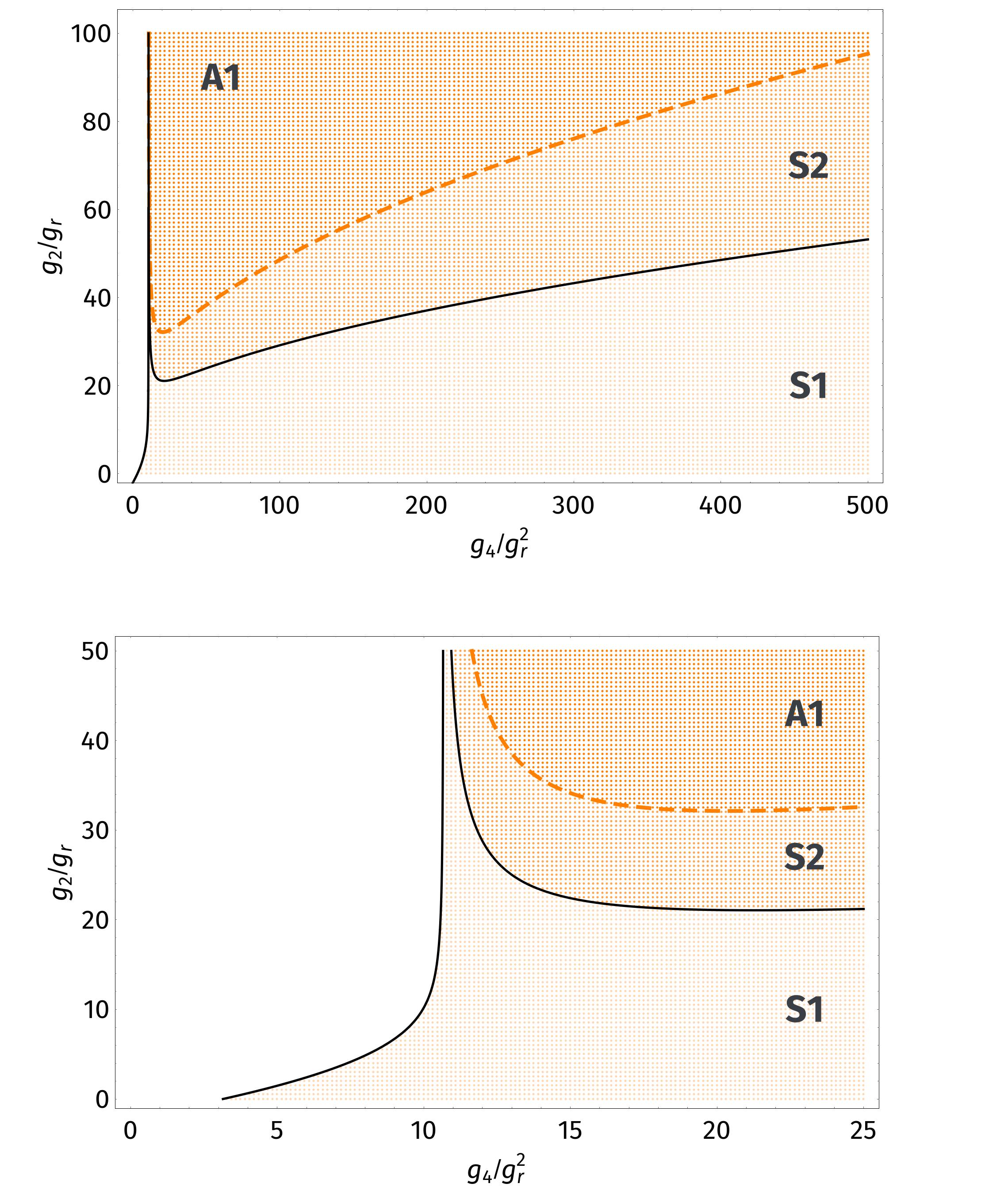}
    \caption{
    Phase diagram of the second-order action \eqref{Seff} for ${g_r = 0.02}$, with a zoomed-in view shown in the lower panel.
    Densely distributed dots in varying shades of orange represent numerically confirmed S1, S2 and would-be A1-phase solutions.
    The solid black line shows the exact S1-phase boundary, as given by \eqref{g22order} and \eqref{S1-existance-exact}. The dashed orange line indicates the exact S2/A1 transition line \eqref{S2/A1-exact}, which is not realized in this model, as the A1 phase is energetically disfavored compared to the existing S2 phase.}
    \label{figure:07}
\end{figure}

An inspection of the phase diagram in Figure \ref{figure:07} reveals a region near the $g_2$ axis  where our system of equations for the eigenvalue distribution lacks solutions. This “void” indicates a domain that is inaccessible within our multitrace model framework. It would be worthwhile to investigate whether HMC simulations applied to the multitrace action yield meaningful configurations in this region. 

As one moves along the positive $g_4$ axis, this void transitions into a region supporting stable S1 solutions for lower values of $g_2$. For higher values of $g_2$, the “void” similarly gives way to an S1 region, which eventually evolves into the S2 phase along the positive $g_4$ direction.

The absence of solutions in the “void” region can be understood by examining the condition for the existence of the S1 solution, specifically, the one arising from the $z^{-1}$-coefficient. For the second-order effective action, this yields the following equation for the support radius $\sqrt{\delta}$:
\begin{align}
\frac{\delta}{12 \left(3-8 g_r^2 \delta ^2 \right)}\cdot \left(-18 g_2+27 g_4\delta+144 g_r-288 g_r^2\delta +24 g_4 g_r^2\delta ^3 -256 g_r^4\delta ^3 \right)=1 \, .
\end{align} 
Solutions to this equation involve several nested square roots. Imposing that the arguments of these roots remain non-negative places constraints on the parameters $g_r$, $g_4$, and $g_2$. The locus where these square roots vanish defines the existence line—the boundary beyond which S1 solutions cease to exist in the phase diagram. For the second-order effective action, this boundary is given explicitly by:
\begin{equation}\label{S1-existance-exact}
    g_2= - \sqrt{12} \sqrt{-g_4 + \dfrac{32}{3} g_r^2} + 8 g_r +
    \frac{\dfrac{64}{9\sqrt{3}} g_r^2}{\sqrt{-g_4 + \dfrac{32}{3} g_r^2}} \, ,
    \qquad g_4 < \dfrac{32}{3} g_r^2 \, .
\end{equation}

Finally, numerical solutions of the aforementioned self-consistent system yield a detailed phase diagram for the second-order\footnote{The numerical determination of the support parameters for the higher-order effective action proved infeasible due to the increasing complexity of the resulting equations, which rendered the computations intractable.} effective action (Figure \ref{figure:07}), in excellent agreement with the analytically derived exact transition lines and phase boundaries.

\section{Monte Carlo Simulation of Full Model $S$}
\label{appendix:extrapolation}

An alternative approach to determining the eigenvalue distributions is to perform Monte Carlo simulations. In particular, we have applied HMC methods \cite{betancourt2018conceptual} enhanced with an eigenvalue-flipping algorithm \cite{Kovacik:2022kfh} during the thermalization stage. This combined approach provides nonperturbative insights and is especially valuable for exploring phase transitions in regimes where our analytical approximations break down, i.e. in the $g_4 \to 0$ limit.

To identify phase transitions from the resulting eigenvalue distributions, we focus on the hallmark of the S1/S2 transition: the splitting of the eigenvalue distribution into two distinct cuts as the system approaches the critical point. In practice, we monitor the midpoint of the eigenvalue distribution moving toward zero as an indicator of this split. Finite-$N$ effects tend to smooth out the emergence of two cuts, so we analyze the distribution behavior over a range of system parameters to pinpoint the transition more reliably. The eigenvalue distributions themselves are obtained from eigenvalue histograms for the field configurations generated by the HMC simulation.

\begin{figure}[t]
\centering  
\includegraphics[width=0.85\textwidth]{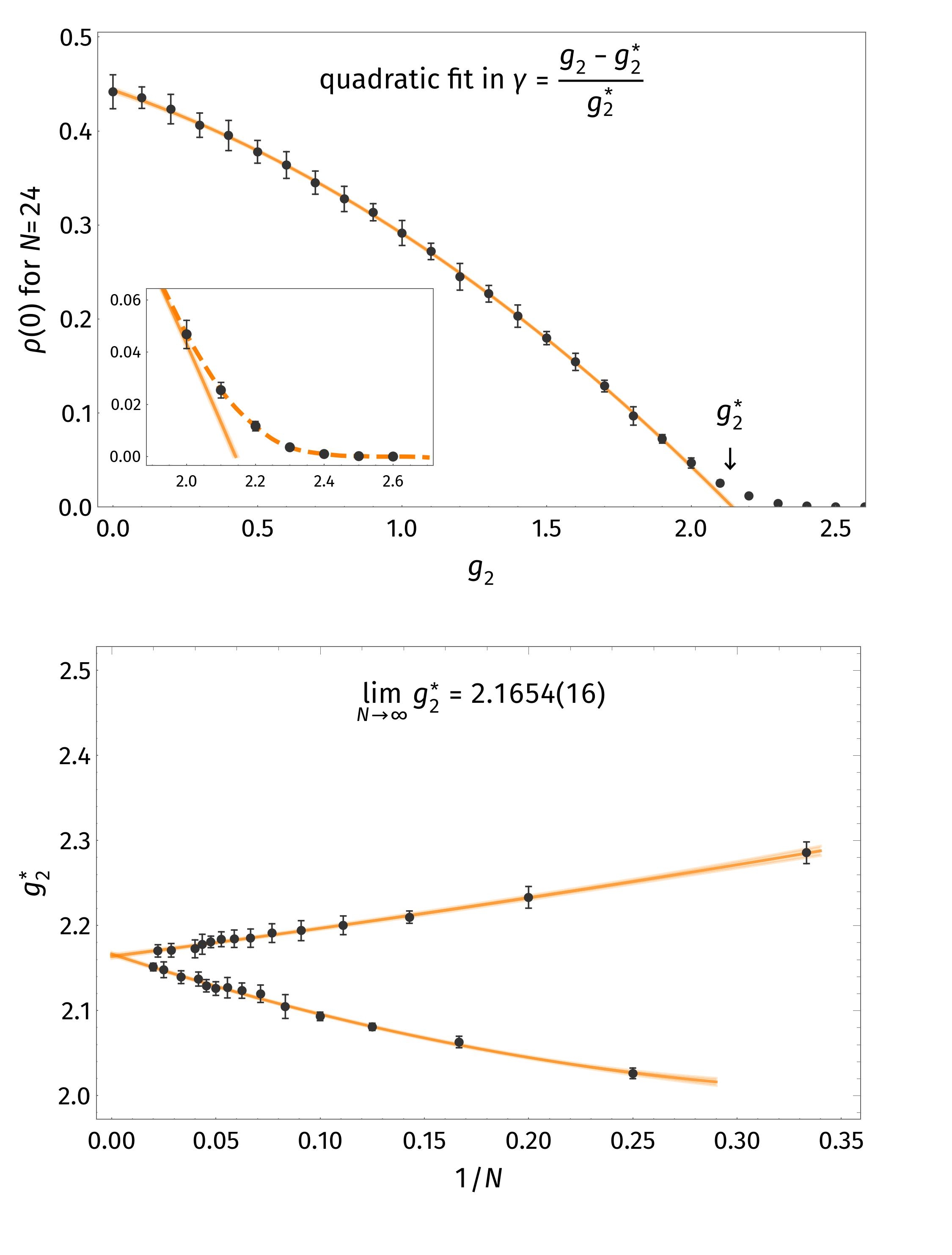}
\caption{
The top panel illustrates how the transition point $g_2^*$ is extracted from $\rho(0)$ data at fixed $N=24$ (for $(g_4,\,g_r)=(1,\,0.02)$) by varying $g_2$. The bottom panel shows the extrapolation of the resulting $g_2^*$ values to the large-$N$ limit. Note that data points for even and odd $N$ fall on separate branches of the curve. Error bars have been magnified by a factor of 2.5 for better visibility.
}
\label{figure:08}
\end{figure}

As observed in our previous work \cite{Prekrat:2022sir}, the midpoint of the eigenvalue distribution decreases approximately linearly with the parameter
\begin{equation}
    \gamma = \frac{g_2 - g_2^*}{g_2^*}\, ,
\end{equation}
as the system approaches the transition point $g_2^*$. In the present study, we extend those findings by incorporating $O(g_r^4)$ contributions to assess the convergence of our numerical results with the analytic prediction. The updated expression for the eigenvalue density at zero (the center of the distribution) is given by
\begin{align}\label{rho0-expanded-v2}
    \rho(0)  = -\frac{\sqrt[4]{4g_4}}{\pi}
    \bigg(
    1 + \epsilon - \frac{1}{6}\epsilon^2 &- \frac{1}{2}\epsilon^3 + \frac{19}{120}\epsilon^4
    \bigg)
    \,\cdot\,
    \notag \\
    \,\cdot\,
    \Bigg\{
    \gamma
    & +
    \frac{1}{8}
    \lr{1 + \epsilon - \frac{8}{3}\epsilon^2 - 3\epsilon^3 + \frac{16}{5}\epsilon^4}
    \gamma^2-
    \notag \\
    & -
    \frac{1}{6}\epsilon^2 
    \lr{1 + 2\epsilon - \frac{7}{3}\epsilon^2}
    \gamma^3-
    \notag \\
    & -
    \frac{1}{512}
    \lr{1 + 3\epsilon + \frac{49}{3}\epsilon^2 + 40\epsilon^3 - \frac{722}{5}\epsilon^4}
    \gamma^4
   \Bigg\}\, .
\end{align}
Figure \ref{figure:08} (top panel) illustrates this behavior for ${N=24}$. The \enquote{tail} visible in the zoomed-in region of that plot is attributed to finite-$N$ effects. We carried out the procedure of fitting the quadratic function in $\gamma$ (since higher-order terms in \eqref{rho0-expanded-v2} are strongly suppressed) to $\rho(0)$ data by incrementally including additional data points. The process was halted once the residuals of the newly added points became significantly larger than typical values and displayed a systematic positive trend—indicating the onset of the tail.


We have then compiled estimates of the transition point $g_2^*$, the linear coefficient\footnote{We use the symbol $\#$ to denote the coefficient of a term, so $\#\gamma$ refers to the coefficient multiplying $\gamma$.} $\#\gamma$
\begin{equation}
    \#\gamma
    \;\leftrightarrow\; 
    \frac{\sqrt[4]{4g_4}}{\pi}
    \lr{
    1 + \epsilon - \frac{1}{6}\epsilon^2 - \frac{1}{2}\epsilon^3 + \frac{19}{120}\epsilon^4
    }\, ,
\end{equation}
and the ratio of the quadratic to linear coefficients $\#\gamma^2/\#\gamma$
\begin{equation}
    \frac{\,\#\gamma^2}{\,\#\gamma^{\phantom{2}}} 
    \;\leftrightarrow\;  
    \frac{1}{8}
    \lr{1 + \epsilon - \frac{8}{3}\epsilon^2 - 3\epsilon^3 + \frac{16}{5}\epsilon^4}\, ,
\end{equation}
 for various values of $N$. To extrapolate these quantities to the ${N \to \infty}$ limit, we performed polynomial fits in $1/N$, using at most quadratic terms. Notably, data points corresponding to even and odd values of $N$ fall on distinct branches in the bottom plot of Figure \ref{figure:08}. This bifurcation arises because odd $N$ values exhibit a small local maximum in the eigenvalue density $\rho$ at ${\lambda=0}$, necessitating separate treatment. Our dataset includes matrix sizes up to ${N=50}$. Remarkably, even the use of smaller matrix sizes gives consistent albeit less precise $N \to \infty$ results. 

To facilitate a comparison between analytical and numerical results, we require error estimates for our analytical predictions. Examining the coefficients in equations \eqref{S1-S2-line-v2} and \eqref{rho0-expanded-v2}, we estimate the uncertainties by assuming that these series converge and that the expansion coefficients are of order $O(10^{-1})$ for $g_2^*$ and $\#\gamma$, and $O(10^0)$ for $\#\gamma^2 / \#\gamma$. Truncating the series at order $O(g_r^{\tau})$ yields the following error estimates:
\begin{equation}
    \Delta g_2^* \sim
    2\sqrt{g_4} \sum_{n = \tau}^\infty 10^{-1} \cdot \epsilon^n 
    = \frac{2\sqrt{g_4}}{10} \cdot
    \frac{\epsilon^\tau}{1 - \epsilon} \, ,
\end{equation}
\begin{equation}
    \Delta\#\gamma\sim
    \frac{\sqrt[4]{4g_4}}{\pi}\sum_{n = \tau}^\infty 10^{-1} \cdot \epsilon^n 
    = \frac{\sqrt[4]{4g_4}}{10\pi} \cdot
    \frac{\epsilon^\tau}{1 - \epsilon} \, ,
\end{equation}
\begin{equation}
    \Delta\frac{\,\#\gamma^2}{\,\#\gamma^{\phantom{2}}} \sim 
    \frac{1}{8} \sum_{n = \tau}^\infty 10^0 \cdot \epsilon^n 
    = \frac{1}{8} \cdot
    \frac{\epsilon^\tau}{1 - \epsilon} \, .
\end{equation}
Numerical values corresponding to these error estimates are presented in Tables \ref{tab1}--\ref{tab3}, alongside an order-by-order convergence analysis of the transition point. This analysis clarifies the previously observed discrepancy in the $\#\gamma^2 / \#\gamma$ results reported in \cite{Prekrat:2022sir}. Specifically, for the value ${g_r = 0.1}$ used in that study, the convergence of the analytical series was insufficiently rapid, leading to a significant error at the truncation order employed.

\begin{table}[t]    
    \centering
    \begin{tabular}{c|cccc|c}
         $g_2^*$ & $O(g_r^0)$ & $O(g_r^2)$ & $O(g_r^4)$ & $O(g_r^6)$ & $\Delta \approx$
         \\ \hline
         $g_4 = 1$, $g_r = 0.02$ & 2.000000 & 2.160000 & 2.164267 & \underline{2.164269} & \underline{$5 \cdot 10^{-9}$}
         \\ 
         $g_4 = 1$, $g_r = 0.1\phantom{0}$ & 2.000000 & 2.800000 & \underline{2.906667} & 2.908373 & \underline{$4 \cdot 10^{-3}$} 
     \end{tabular}
    \caption{
    Convergence analysis for $g_2^*$. Each error $\Delta$ corresponds to the estimated uncertainty at the underlined order of approximation, which denotes the highest order reached in the simulation for the specified coupling parameters.}
    \label{tab1}
\end{table}

\begin{table}[t]     
    \centering
    \begin{tabular}{c|ccccc|c}
         $\#\gamma$ & $O(g_r^0)$ & $O(g_r^1)$ & $O(g_r^2)$ & $O(g_r^3)$ & $O(g_r^4)$ & $\Delta \approx$ \\
         \hline
         $g_4 = 1$, $g_r = 0.02$ & 0.45016 & 0.48617 & 0.48569 & 0.48558 & \underline{0.48558} & \underline{$2 \cdot 10^{-7}$} \\ 
         $g_4 = 1$, $g_r = 0.1\phantom{0}$ & 0.45016 & 0.63022 & \underline{0.61822} & 0.60381 & 0.60564 & \underline{$5 \cdot 10^{-3}$} \\ 
     \end{tabular}
    \caption{Convergence analysis for $\#\gamma$. Each error $\Delta$ corresponds to the estimated uncertainty at the underlined order of approximation, which denotes the highest order reached in the simulation for the specified coupling parameters.}
    \label{tab2}
\end{table}

\begin{table}[t]     
    \centering
    \begin{tabular}{c|ccccc|c}
         $\#\gamma^2 / \#\gamma$ & $O(g_r^0)$ & $O(g_r^1)$ & $O(g_r^2)$ & $O(g_r^3)$ & $O(g_r^4)$ & $\Delta \approx$ \\
         \hline
         $g_4 = 1$, $g_r = 0.02$ & 0.12500 & 0.13500 & 0.13287 & 0.13267 & \underline{0.13269} & \underline{$5 \cdot 10^{-7}$} \\ 
         $g_4 = 1$, $g_r = 0.1\phantom{0}$ & 0.12500 & 0.17500 & \underline{0.12167} & 0.09767 & 0.10791 & \underline{$2 \cdot 10^{-2}$} \\ 
     \end{tabular}
    \caption{Convergence analysis for $\#\gamma^2 / \#\gamma$. Each error $\Delta$ corresponds to the estimated uncertainty at the underlined order of approximation, which denotes the highest order reached in the simulation for the specified coupling parameters.}
    \label{tab3}
\end{table}

We have here obtained high-precision numerical results in a regime where analytical expression \eqref{S1-S2-line-v2} shows signs of convergence already at order $O(g_r^6)$, specifically: \begin{equation}
    \epsilon \ll 1
    \quad\Rightarrow\quad
    g_4 \gg 16g_r^2 \, .
\end{equation}
As we can see, the analytical prediction \eqref{S1-S2-line} aligns exceptionally well with numerical result for the model parameters ${(g_r,\,g_4) = (0.02, \,1)}$:
\begin{subequations}
\begin{align}
\text{analytical:} \qquad  g_2^* &\approx 2.164269385(5) \, ,
\\
\text{numerical:} \qquad  g_2^* &\approx 2.1654(16) \, .
\end{align}
\end{subequations}
Furthermore, the coefficients of $\gamma$ and $\gamma^2$ terms closely match the predictions of \eqref{rho0-expanded-v2}:
\begin{subequations}
\begin{align}
\text{analytical:} \qquad \rho(0) &\approx -0.4855783(2)\cdot\lr{\gamma + 0.1326911(5)\cdot\gamma^2} \, ,
\\
\text{numerical:} \qquad \rho(0) &\approx -0.4849(8)\cdot\lr{\gamma + 0.1312(16)\cdot\gamma^2} \, .
\end{align}
\end{subequations}
A more detailed analysis of curvature effects, presented in Table \ref{tab:curvature-effects}, reveals that its contributions to $g_2^*$ and $\#\gamma$ are confirmed within a 1--2 percent uncertainty. Even the subtler effect in $\#\gamma^2 / \#\gamma$ is corroborated at approximately the $4\sigma$ level.

\begin{table}[t]   
    \centering
    \begin{tabular}{c|ccc}
         contribution to & $g_2^*$ & $\#\gamma$ & $\#\gamma^2 / \#\gamma$ \\
         \hline
         analytical & 0.164269385(5) & 0.0354201(2) & 0.0076911(5) \\ 
         numerical &  0.1654(16) & 0.0348(8) & 0.0062(16) \\ 
     \end{tabular}
    \caption{Curvature contribution to $g_2^*$, the linear coefficient $\#\gamma$ and the ratio $\#\gamma^2 / \#\gamma$ for ${g_4 = 1}$ and ${g_r = 0.02}$.}
    \label{tab:curvature-effects}
\end{table}

\begin{figure}[t]
\centering  
\includegraphics[width=0.8\textwidth]{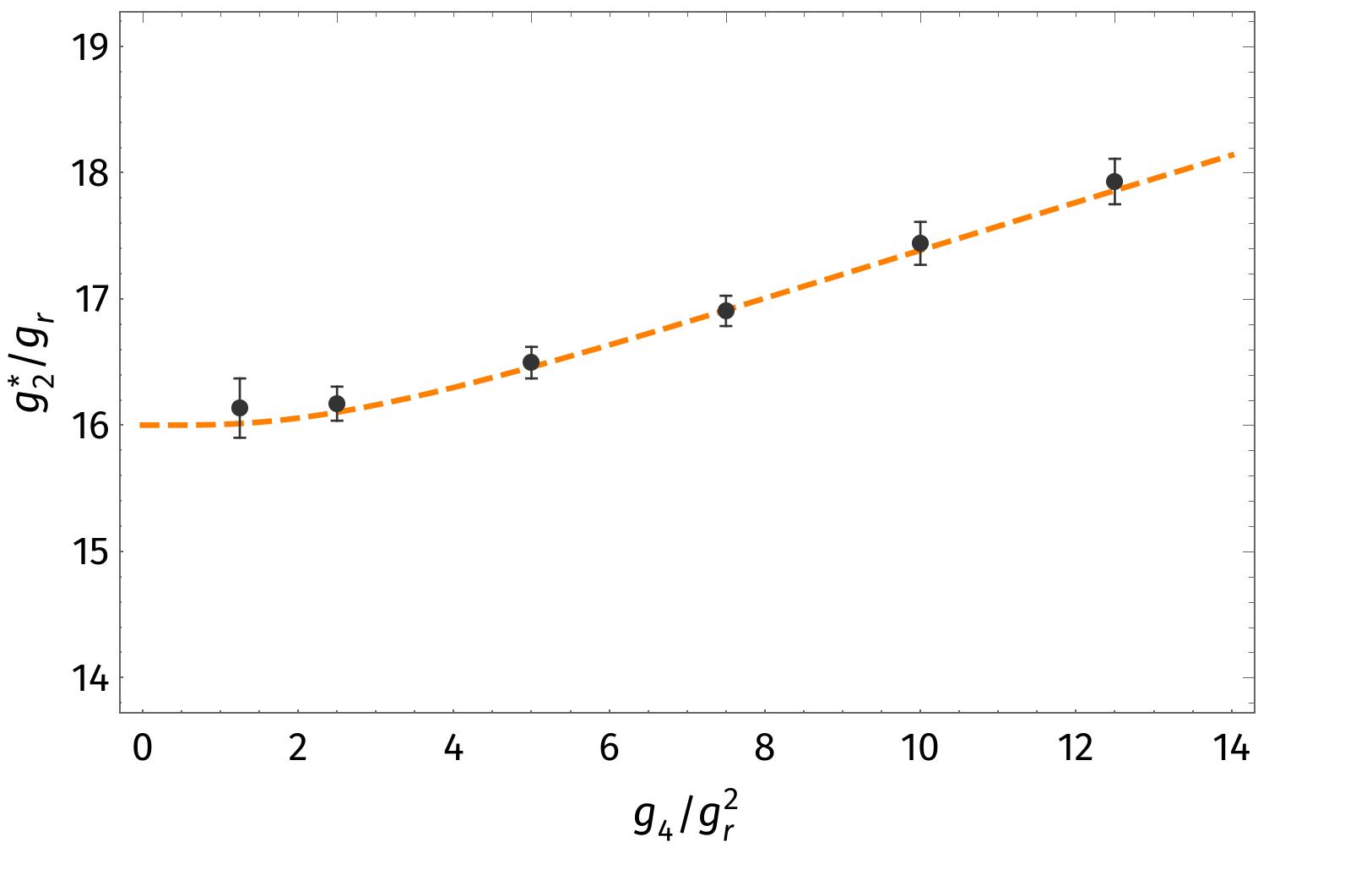}
\includegraphics[width=0.8\textwidth]{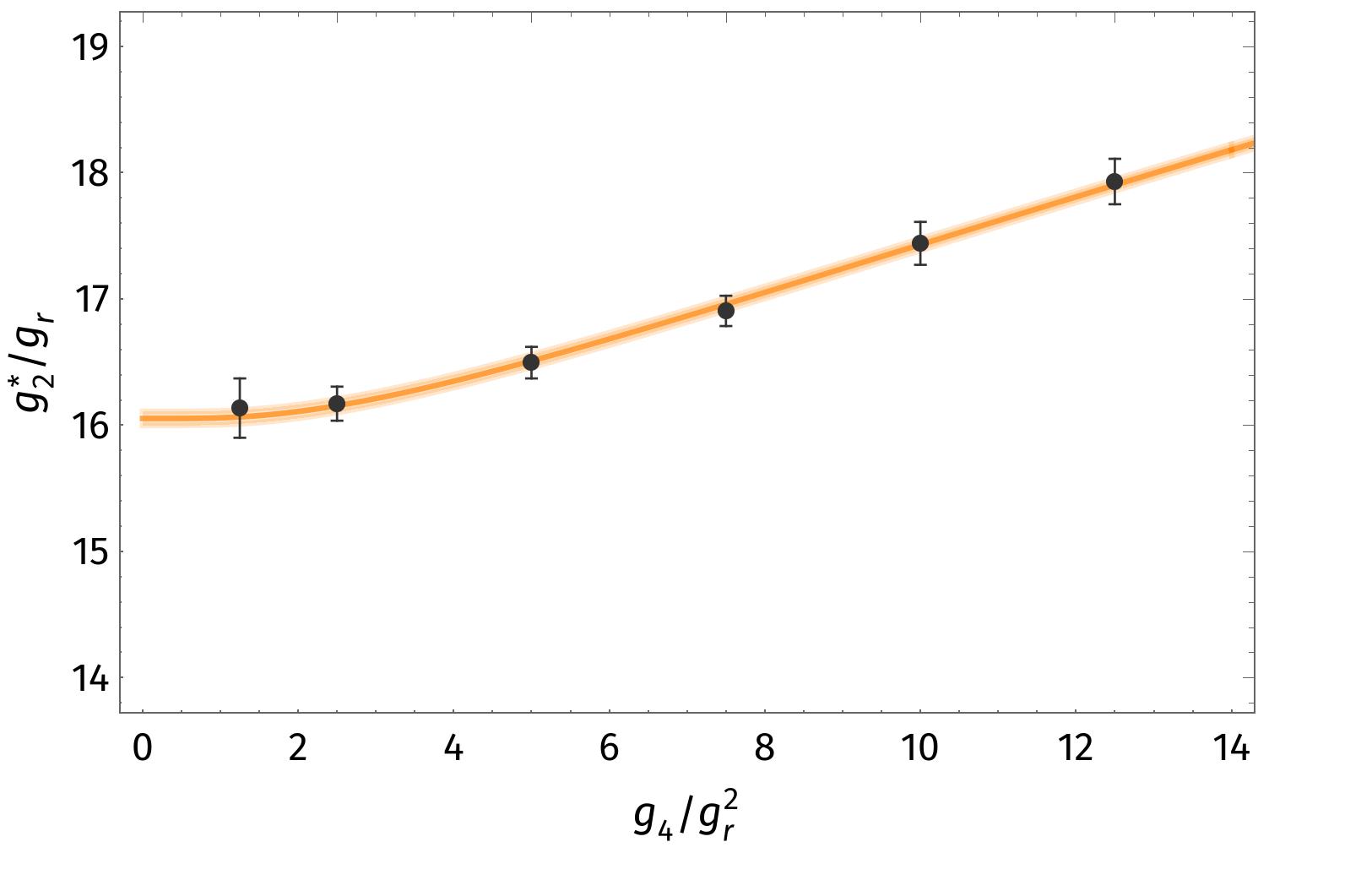}
\caption{
Onset of the S1/S2 transition line for ${g_r = 0.02}$. The top panel shows the proposed functional form from \eqref{old-guess}. The bottom panel presents a fit of the same form with the coefficient $8g_r$ treated as a free parameter. The fit was performed using large-$N$ extrapolated data obtained via HMC simulation of action \eqref{GW-without-K} (additional data points lie outside the zoomed-in region). The resulting numerical estimate, ${g_2^*(0) = 16.05(6)g_r}$ is in excellent agreement with the expected shift of $16 g_r$.
}
\label{figure:09}
\end{figure}

We have also investigated the regime where ${g_4 = O(g_r^2)}$, domain in which our perturbative results are no longer reliable. This regime is particularly significant when considering the shift of the triple point. In our previous study \cite{Prekrat:2022sir}, we proposed the ansatz:
\begin{equation}\label{old-guess}
    g_2^* = \dfrac{16 g_r}{1- \exp \lr{-\dfrac{8g_r}{\sqrt{g_4}}}} 
    = 8 g_r \lr{1 + \coth \dfrac{8g_r}{2\sqrt{g_4}}} 
\end{equation}
which shares the same $O(g_r^2)$ expansion as our transition line and provides a good fit to the numerical data. Utilizing this functional form, we performed a non-perturbative fit to our large-$N$ limit data points, as depicted in Figure \ref{figure:09}, yielding:
\begin{equation}
    g_2^* = \alpha g_r \lr{1 + \coth \dfrac{\alpha g_r}{2\sqrt{g_4}}},
    \qquad
    \alpha = 8.03(3) \, .
\end{equation}
This result implies that the S1/S2 transition line begins at ${g_4 = 0}$ and ${g_2 = 16.05(6) \, g_r}$, consistent with the expected shift of $16 g_r$ from the origin. An alternative fit, which reproduces the correct $O(g_r^6)$ structure but relies on a speculative extrapolation into the weak self-interaction regime, is presented in the Appendix.

\begin{figure}[t]
\centering  
\includegraphics[width=0.99\textwidth]{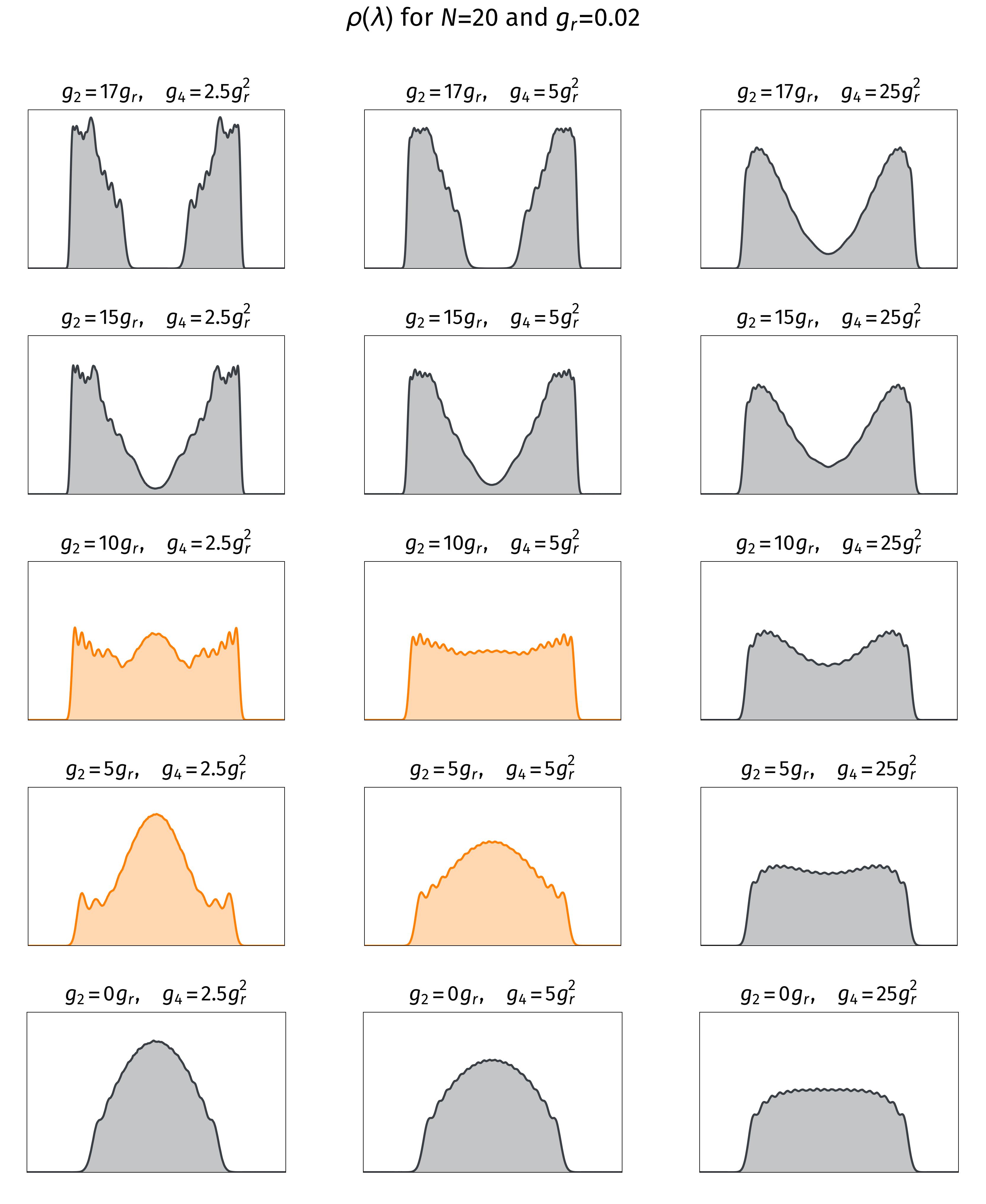}
\caption{
Eigenvalue distributions for ${N = 20}$ and ${g_r = 0.02}$ in the low-$g_4$ regime, illustrating the transition from the S1 to S2 phase (top left corner), which at ${g_4 = 0}$ is expected to occur at ${g_2 = 16g_r}$. Here, $g_4$ increases from left to right and $g_2$ from bottom to top. The orange curves, featuring both central peaks and sharp peripheral eigenvalues, correspond to quantum solutions of the $\Psi_{\!R}^\pm$ type. This deviation from the typical two-cut structure (see Figure~\ref{figure:01}) may signal the emergence of a novel phase.}
\label{figure:10}
\end{figure}

\begin{figure}[t]
\centering  
\includegraphics[width=0.8\textwidth]{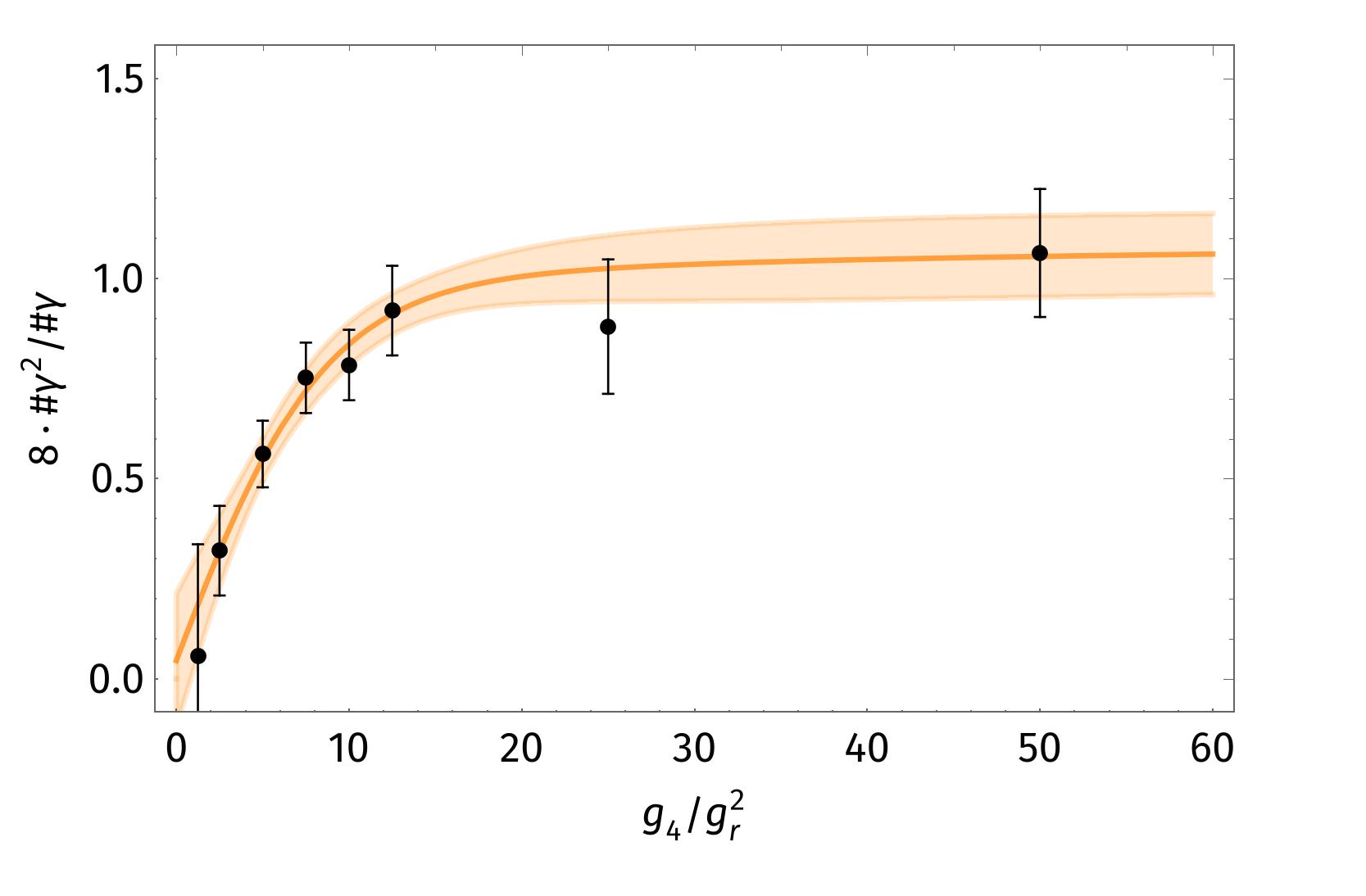}
\caption{
Behavior of the coefficient ratio $\#\gamma^2 / \#\gamma$ at low $g_4$, exhibiting a sharp drop from its asymptotic value of $1/8$ toward zero. This signal correlates with the emergence of central peaks in the eigenvalue distribution (Figure \ref{figure:10}), although no corresponding anomaly was observed in thermodynamic observables.}
\label{figure:11}
\end{figure}

While examining the model’s behavior at low $g_4$, we encountered a novel feature in the eigenvalue distributions. Instead of the well-known pattern as in Figure \ref{figure:01}, the distributions began to exhibit central peaks, with eigenvalues gradually chipping away from the edges of the bulk (see the orange curves in Figure \ref{figure:10}). A corresponding signal was also
observed in the coefficient ratio $\#\gamma^2 / \#\gamma$ at low $g_4$, where this quantity departs sharply from its asymptotic value of $1/8$ and
decreases toward $0$ (Figure \ref{figure:11}). 
Despite these clear structural changes in the eigenvalue distributions,
we did not observe any accompanying qualitative change in standard
thermodynamic observables, such as the free energy, specific heat,
magnetization, susceptibility, or Binder cumulant, until much deeper
into the weak self-interaction regime. Similar effects were reported
previously in \cite{Prekrat:2020ptq}, and even in simulations of the full
GW model including the kinetic term \cite{Prekrat:2023thesis}, where a
change in susceptibility was observed. However, in those studies the
phenomenon was not investigated systematically, and indications were
reported that the anomalous region shrinks with increasing $N$,
suggesting that it may represent a finite-$N$ artifact rather than a
bona fide phase transition.

Let us now attempt to explain this effect. Consider the exact solution $\Phi_{\!R}$ given by \eqref{phiR}, which satisfies the EOM in the absence of the kinetic term. This solution is well-defined only for ${g_2 \geq 16 g_r}$, so it cannot directly account for behavior at smaller values of $g_2$. However, due to the diagonality of the curvature $R$, the EOM can be solved independently for each eigenvalue by either 0 or $\pm(\Phi_{\!R})_{nn}$.  In other words, it is possible to replace some of the eigenvalues of $\Phi_{\!R}$ by zero while still satisfying the EOM. The simplest such partially-degenerate solution is
\begin{equation}
    (\Psi_{\!R})_{nn} = 
    \begin{cases}
        (\Phi_{\!R})_{nn} & \text{for } n \leq M\, ,
        \\
         0 & \text{for } n > M\, ,
    \end{cases}
\end{equation}
which is now valid for ${g_2 \geq 16g_r M/N}$.

We begin by analyzing the Vandermonde (eigenvalue repulsion) contribution to the effective action for $\Phi_{\!R}$ in $g_2 \geq 16 g_r$ regime. We consider both the A1 version, in which all eigenvalues have the same sign, and the M2 version, denoted $\Phi_{\!R}^\pm$, where signs alternate:
\begin{equation}
    (\Phi_{\!R}^\pm)_{nn}^{\phantom{\pm}} = (-1)^{n}(\Phi_{\!R}^{\phantom{\pm}}\!)_{nn}^{\phantom{\pm}} \ .
\end{equation}

This repulsive term is what prevents the eigenvalues from collapsing onto a single value: instead of sharply defined clusters of degenerate eigenvalues (as one might expect from the naive solutions \eqref{vacuum solutions} with identical eigenvalues), one obtains broadened peaks (as seen in Figure \ref{figure:01}). Since there is no eigenvalue degeneracy in $\Phi_{\!R}$, the repulsion is much weaker in this case. 

Assuming small $g_r$, we approximate $\Phi_{\!R}$ to first order in $g_r$ as
\begin{equation}
    \Phi_{\!R} = \sqrt{\frac{g_2 \1 + g_r R}{2 g_4}} 
    = 
    \sqrt{\frac{g_2\mathrlap{\phantom{R}}}{2 g_4}} \sqrt{\1 + \frac{g_r R}{g_2}}
    \approx
    \sqrt{\frac{g_2\mathrlap{\phantom{R}}}{2 g_4}} \lr{\1 + \frac{g_r R}{2g_2}} .
\end{equation}
This yields the Vandermonde determinant \eqref{vandermonde}: 
\begin{equation}
    \Delta(\Phi_{\!R}) = \lr{-\frac{8g_r}{N\sqrt{2 g_2 g_4}}}^{\!\!\frac{N(N-1)}{2}}\prod_{n=1}^{N-1} n! \, ,
\end{equation}
so the corresponding contribution to the effective action is: 
\begin{align}\label{VandemondePhiR}
    S_\text{eff}^{\Delta}(\Phi_{\!R}) 
    &= \notag
    - N(N-1) \ln \frac{8g_r}{N\sqrt{2 g_2 g_4}}
    - 2 \sum_{n=1}^{N-1} \ln n! \sim
    \\ \notag
    &\sim N^2 \ln N - N^2 \ln \frac{8g_r}{\sqrt{2 g_2 g_4}} - \lr{N^2\ln N - \frac32 N^2} =
    \\ 
    &= \lr{\frac32 - \ln \frac{8g_r}{\sqrt{2 g_2 g_4}}} N^2 \, ,
\end{align}
where we used Stirling’s approximation and the Euler–Maclaurin formula.
For $\Phi_{\!R}^\pm$, we compute the contributions to $\Delta$ from eigenvalue differences between same-sign and opposite-sign pairs separately, and find:
\begin{equation}
    S_\text{eff}^{\Delta} (\Phi_{\!R}^{\pm}) = S_\text{eff}^{\Delta} (\Phi_{\!R}^{\phantom{\pm}}\!) - N^2 \ln 2 \ .
\end{equation}

Meanwhile, the remaining trace part of the effective action \eqref{Seff-v2} estimated up to $O(g_r^2)$, gives:
\begin{equation}
    S_\text{eff}^{\,\tr} (\Phi_{\!R}^{\mathrlap{\phantom{\pm}}}) = 
    S_\text{eff}^{\,\tr} (\Phi_{\!R}^\pm) =
    -\frac{(g_2 - 8g_r)^2 - \frac{1}{3} \cdot (8g_r)^2}{4g_4}N^2 .
\end{equation}
Combining these results, we see that $\Phi_{\!R}^{\pm}$ has lower free energy than $\Phi_{\!R}^{\phantom{\pm}}$, indicating that the A1 phase is not realized. 

In the ${g_4 \to 0}$ limit, the Vandermonde term becomes negligible compared to the rest of the action:
\begin{equation}\label{multicut-at-zero}
    \frac{S_\text{eff}^{\Delta} (\Phi_{\!R}^{\pm})}{S_\text{eff}^{\,\tr} (\Phi_{\!R}^\pm)} 
    \propto 
    -g_4\ln g_4 \to 0 \ ,
    \qquad
    g_4 \to 0 \ .
\end{equation}
Thus, we expect the classical solution $\Phi_{\!R}^{\pm}$ to dominate over the deformations induced by the Vandermonde term in this limit. Conversely, for larger values of $g_4$, the eigenvalues of  $\Phi_{\!R}^{\pm}$ move closer together, the Vandermonde repulsion becomes significant, and the resulting distribution (the right column of Figure \ref{figure:10}) more closely resembles the vacuum solutions in \eqref{vacuum solutions} (Figure \ref{figure:01}).

Although the eigenvalue distribution of $\Phi_{\!R}^{\pm}$ is not strictly symmetric—unlike those obtained from HMC simulations—it becomes approximately symmetric after a small shift:
\begin{equation}
    \Phi_{\!R}^{\pm} 
    \; \longrightarrow \;
    \Phi_{\!R}^{\pm} - \frac{8g_r}{N} \ .
\end{equation}
In simulations, this shift likely arises due to the permutation symmetry of the Vandermonde term. Moreover, the difference between the absolute values of successive eigenvalues of opposite signs vanishes in the large-$N$ limit.

\begin{figure}[t]
\centering  
\includegraphics[width=0.99\textwidth]{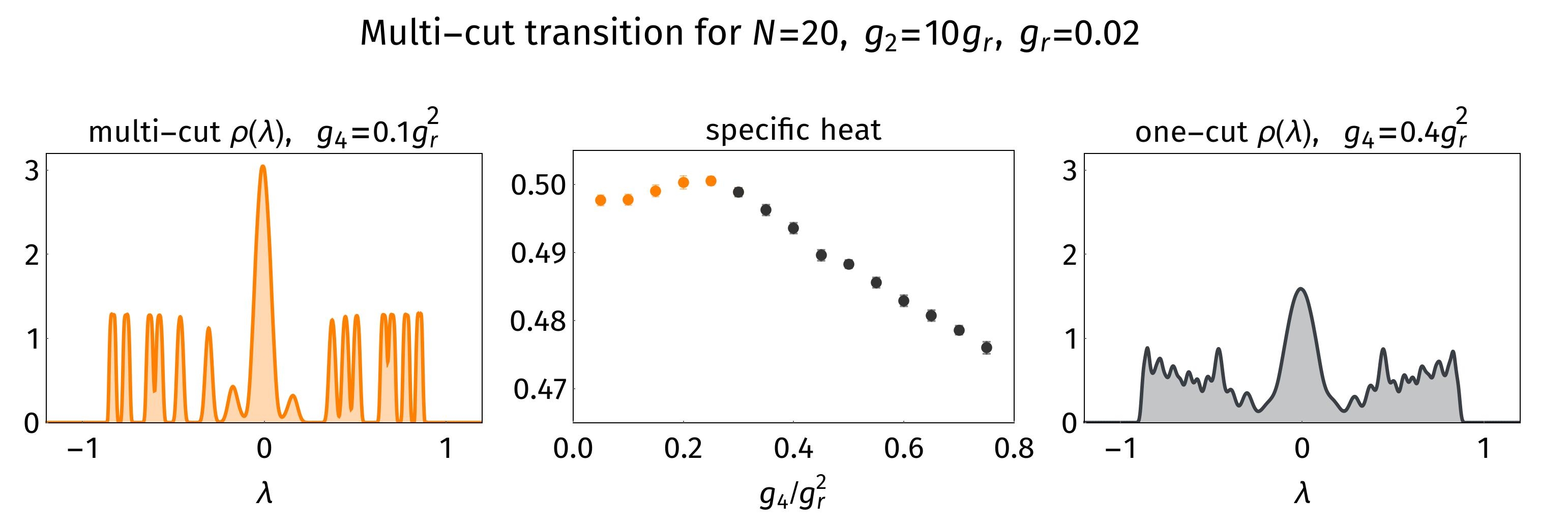}
\caption{
Transition from the one-cut to a multi-cut phase with eigenvalue distribution of the $\Psi_{\!R}^\pm$ form, observed at fixed $N$ and very small ${g_4 = O(10^{-1}g_r^2)}$. The transition is accompanied by a pronounced feature in the profile of specific heat, ${\Var S / N^2}$, consistent with a would-be third-order phase transition. Qualitatively, this behavior is very similar to the well-established third-order transition from the one-cut to the two-cut phase.}
\label{figure:12}
\end{figure}

Now consider the solutions $\Psi_{\!R}^\pm$, obtained by similarly alternating the eigenvalue signs in $\Psi_{\!R}^{\phantom{\pm}\!}$. These solutions remain valid even when $g_2 < 16g_r$. 
To leading order in $g_r$, their zero-eigenvalue sector behaves like a pure potential model in the S1 phase, but with a shifted mass parameter:
\begin{equation}
    g_{2,\text{eff}} = g_2 - 8g_r \ ,
    \qquad
    g_{4,\text{eff}} = g_4  \ .
\end{equation}
In the pure potential model, the S1 eigenvalue density transitions from a central dip to a central peak at $g_2 = -2\sqrt{g_4}$ (unimodal to bimodal distribution), which has the same magnitude but opposite sign as the third-order S1/S2 transition point \cite{Prekrat:2023thesis}. In our case, this condition becomes:
\begin{equation}
    g_{2,\text{eff}} = -2 \sqrt{g_{4,\text{eff}}} \ ,
\end{equation}
which translates into:
\begin{equation}
    g_2 = 8g_r - 2 \sqrt{g_4} \ .
\end{equation}
For this behavior to appear in the $g_2 \geq 0$ region—where $\Psi_{\!R}$ solves the EOM—we must have:
\begin{equation}
    g_4 \leq 16g_r^2 \ ,
\end{equation}
in agreement with the trend-change point in Figure \ref{figure:11}. 
Due to the degeneracy of the zero eigenvalues, they are necessarily smeared and begin to overlap with the non-zero sector of $\Psi_{\!R}^\pm$, effectively pushing the latter toward the edges of the support. This explains both the emergence of a central bulk and the formation of sharp peripheral peaks seen in the orange curves of Figure~\ref{figure:10}. 

Furthermore, in the deep small-$g_4$ regime we observed a transition to a
multi-cut phase of the $\Psi_{\!R}^\pm$ form (Figure~\ref{figure:12}),
consistent with the arguments of \eqref{multicut-at-zero}. This behavior strongly resembles the third-order transition from the S1
to the S2 phase in the pure potential model: in both cases the transition
is characterized by a sharp plateauing of the specific heat even at
finite $N$, mirrored by a change in the topology of the eigenvalue
distribution, and accompanied by a gradual deformation of the
distribution and the development of multiple peaks over a wider region of parameter space.
Whether this effect persists in the large-$N$ limit and develops into a
genuine phase transition remains an open question and is the subject of
ongoing work. Figure~\ref{figure:12} suggests that, if this occurs, the
infinite number of narrow peaks located to the left and right of the
broader central peak would merge, effectively yielding a 3-cut phase.

This possibility offers an alternative way to look at the emergence of the central peak. In addition to the familiar 1-cut and 2-cut solutions, the original pure potential model also admits a 3-cut solution, in which the eigenvalues accumulate around the central maximum of the potential. In the pure potential case, this solution is unstable and never realized, which is why it is rarely discussed in the literature. Nevertheless, it is not difficult to imagine that higher-order terms in the effective action \eqref{Seff-v2} could stabilize this solution by generating a local minimum at the origin through contributions such as $\tr\!\Lambda^8$. Whether such a solution becomes energetically favored then depends on a delicate interplay among the various terms in the effective action and on the free energies of the competing stable configurations for given parameter values. Importantly, this mechanism could persist in the large-$N$ limit, although  a detailed investigation is left for future work.

As an end note, we point out a superficial resemblance to Cooper pair formation: the emerging pairs of eigenvalues with opposite signs can be linked to electron pairs with opposite spin orientations. In this analogy, the disordered S1 phase at \enquote{high temperature} gradually \enquote{condenses} into the ordered S2 phase, reminiscent of the onset of superconductivity. Alternatively, since our model is essentially equivalent to that of \cite{Jeevanesan:2019jty}—where our curvature term can be interpreted as arising from either a magnetic field or system rotation—the eigenvalue behaviour may also be viewed as describing superfluid vortices undergoing a (BKT-like) transition from a crystalline to a fluid phase.

\section{Conclusions \& Outlook}

Our results demonstrate that introducing a curvature term into the $\lambda\phi^4_\star$ model significantly alters its phase structure, most notably by shifting the phase transition lines. 

Although we were unable to provide an analytical prediction for the triple point shift via extrapolation from the weak self-coupling regime, we obtained highly precise results in the strong self-coupling limit. These findings were further supported by HMC simulations, which confirmed the expected magnitude of the shift.

This shift plays a pivotal role in the renormalizability of the GW model, as it suppresses the NC striped phase—known to obstruct renormalization through UV/IR mixing. However, the interplay between this suppression mechanism and the model’s kinetic term remains poorly understood. A complete picture will require disentangling and comparing the respective contributions of the curvature and kinetic terms—an issue currently under investigation.

When viewed in terms of unscaled parameters, the triple point shift effectively removes the striped phase from the physically relevant region. This applies both to the renormalizable GW model and to the large-$N$ limit of the $\lambda\phi^4_\text{\tiny GW}$ model, where the curvature coupling vanishes. In contrast, the original, nonrenormalizable $\lambda\phi^4_\star$ model lacks such a protective mechanism, with its triple point remaining fixed at the origin of the phase diagram.

Interestingly, we also identified a possible novel multi-cut phase governed by vacuum solutions modified by the curvature term. At finite
matrix size, this regime exhibits a sharp plateauing of the specific
heat, closely resembling the well-known third-order S1/S2 transition of the pure potential model. Determining whether this structure survives
the large-$N$ limit and constitutes a genuine thermodynamic phase
transition remains an important open problem that will be addressed in
future work.

We believe that the link between the absence of the striped phase and the emergence of renormalizability has broader significance and may extend to other NC models. A natural next step in testing this hypothesis is the simulation of the related—but nonrenormalizable—NC $U(1)$ gauge model \cite{Buric:2016lly}, which is currently under study. This model features two competing classical vacua: a trivial vacuum and a stripe-like vacuum proportional to the NC coordinates. By identifying which vacuum is energetically preferred, we hope to assess whether the persistence of the striped phase indeed correlates with nonrenormalizability. Insights of this kind could inform the construction of consistent, renormalizable NC gauge models—an essential step toward generalizing the successes of the GW model beyond scalar field theory.

\appendix
\section{Alternative Nonperturbative Ansatz for the Transition Line}

When the S1/S2 transition line \eqref{S1-S2-line-v2} is
expressed as
\begin{equation}
    \frac{g_2 - 8g_r - 2\sqrt{g_4}}{8g_r} =
    \frac{1}{3}\,\epsilon
    + \frac{1}{30}\,\epsilon^3
    - \frac{1}{42}\,\epsilon^5 \, ,
\end{equation}
we observe that the coefficients on the right-hand side correspond to
differences of Bernoulli numbers,
\begin{equation}
    B_1 - B_2 = \frac{1}{3} \, ,
    \qquad
    B_3 - B_4 = \frac{1}{30} \, ,
    \qquad
    B_5 - B_6 = -\frac{1}{42} \, .
\end{equation}
Assuming that this pattern continues, and recalling the asymptotic
expansion of the trigamma function,
\begin{equation}
    \psi_1(z) \sim \sum_{n=0}^\infty \frac{B_n}{z^{n+1}} \, ,
\end{equation}
where all odd-indexed Bernoulli numbers (except $B_1 = 1/2$) vanish, we propose the following candidate expression for the transition line:
\begin{equation}
    \frac{g_2 - 8g_r - 2\sqrt{g_4}}{8g_r}
    =
    \frac{2 + \epsilon + \epsilon^2}{2\epsilon}
    - \frac{\psi_1\!\lr{\epsilon^{-1}}}{\epsilon^2} \, .
\end{equation}

\begin{figure}[t]
\centering
\includegraphics[width=0.8\textwidth]{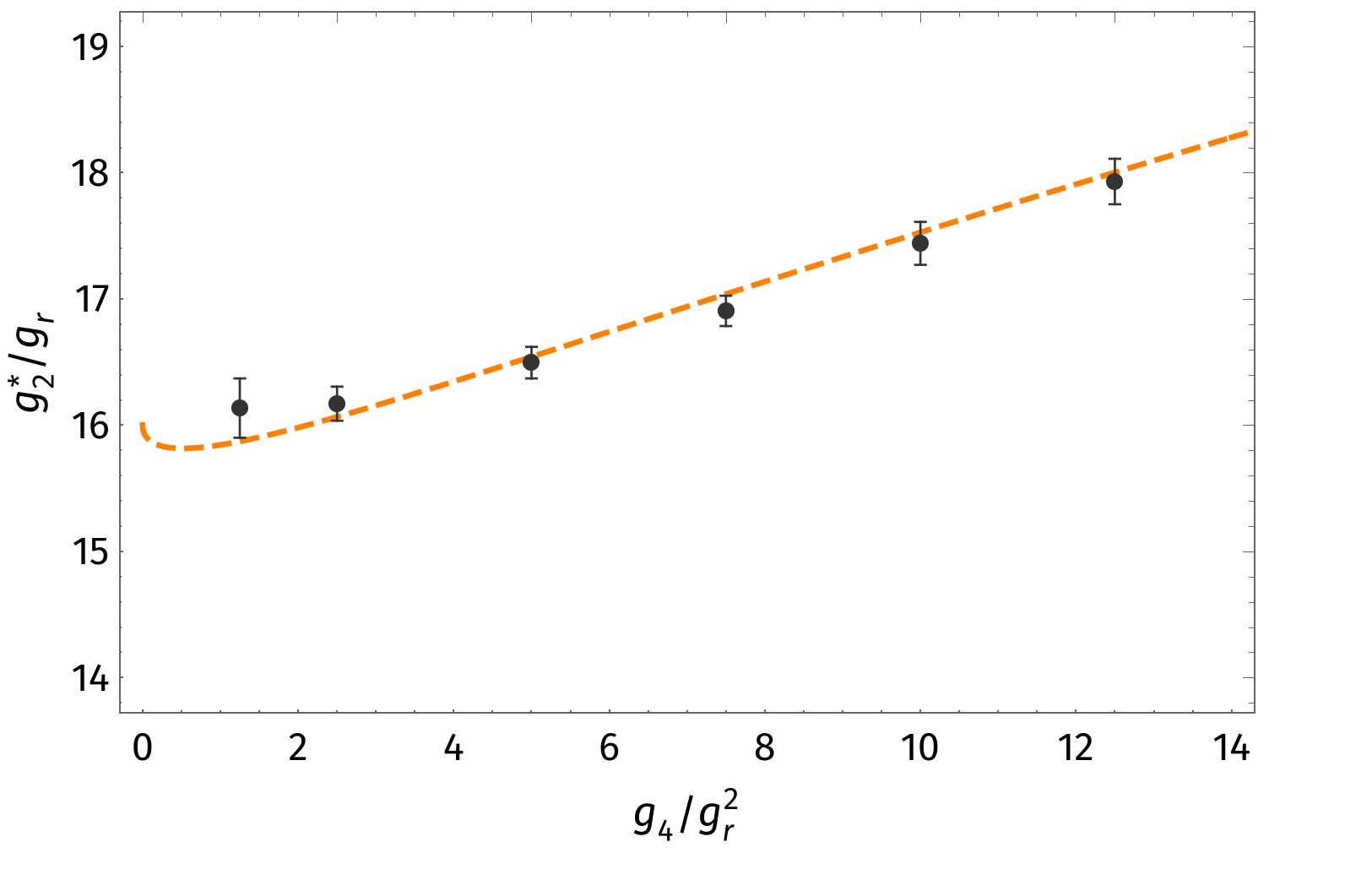}
\includegraphics[width=0.8\textwidth]{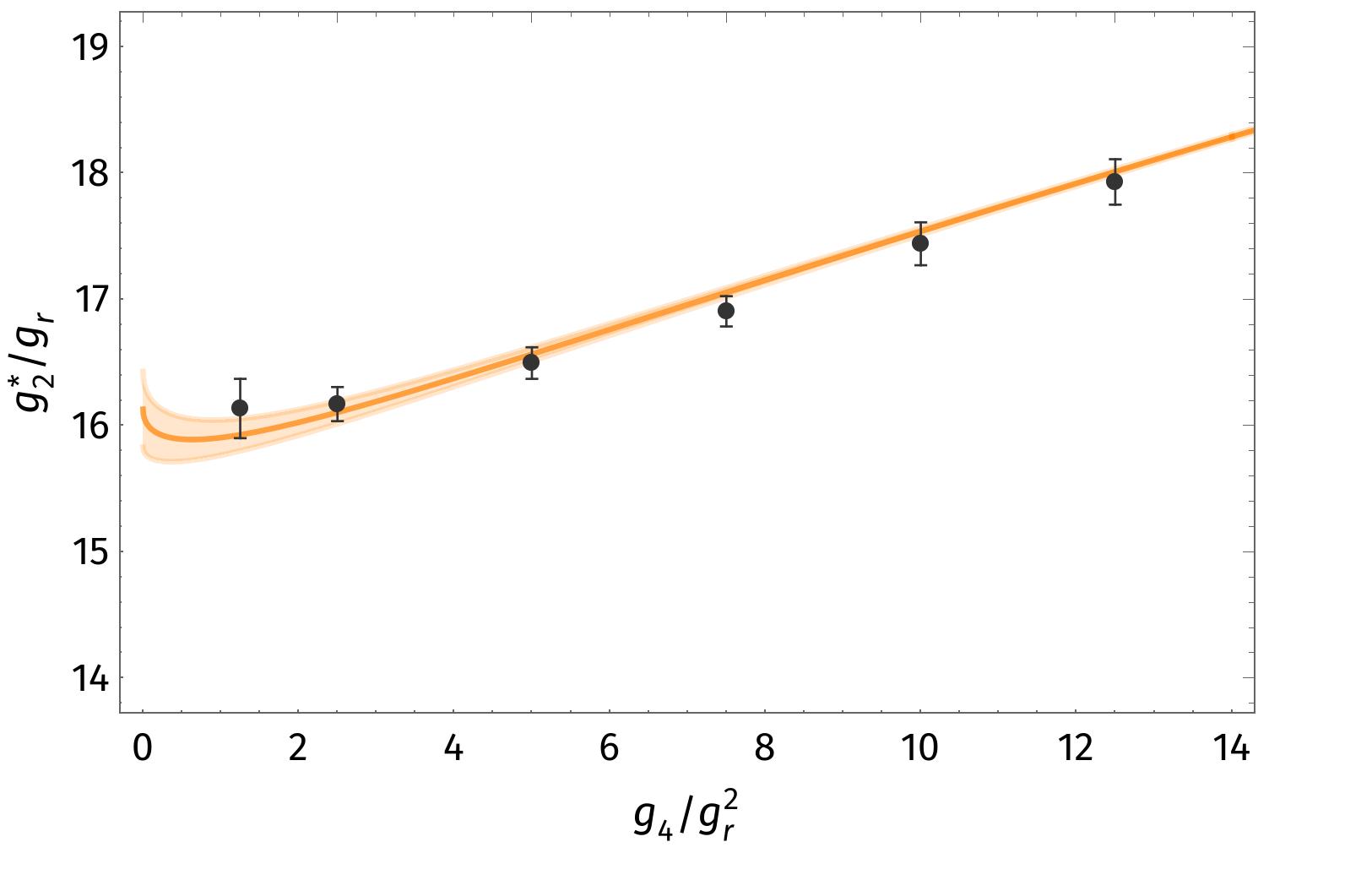}
\caption{
Onset of the S1/S2 transition line for $g_r = 0.02$.
The top panel shows the proposed functional form \eqref{new-guess}.
The bottom panel presents a fit of this form with the exponent $3$ in
\eqref{fix} treated as a free parameter, yielding a fitted value of
$3.03(8)$. The fit was performed using large-$N$–extrapolated data
obtained via HMC simulations of the action \eqref{GW-without-K}
(additional data points lie outside the zoomed-in region).
The resulting estimate, $g_2^*(0) = 16.1(3)\,g_r$, is consistent with
the expected shift of $16 g_r$.
}
\label{figure:13}
\end{figure}

As $g_4 \to \infty$ at fixed $g_r$, $\epsilon \to 0$, and the right-hand
side approaches zero, as expected. However, in the opposite limit
$g_4 \to 0$, the expression diverges, in stark contrast to our Monte
Carlo simulations. The point at which the curve turns toward infinity
occurs around
\begin{equation}
    g_4 \approx 0.4 \times 16 g_r^2 < 7 g_r^2 \, .
\end{equation}
In our simulations, however, we observe no sign of divergence between
the origin and $g_4 = 7 g_r^2$ (see Figure~\ref{figure:13}). This
discrepancy suggests that even if the proposed asymptotic series is
correct, an alternative expression is required—one that shares the same
asymptotic expansion but exhibits the correct behavior at small $g_4$.

Such a modification can be achieved by incorporating an exponentially
suppressed term that has no asymptotic series expansion at infinity.
For example, the desired $16 g_r$ shift can be implemented through the
following modification, which is one of many possible choices but the
simplest we have identified:
\begin{equation}\label{fix}
    \frac{\epsilon}{2}
    \;\longrightarrow\;
    \frac{\epsilon}{2}
    \lr{1 - \exp\lr{-\frac{3}{\epsilon}}} \, ,
\end{equation}
leading to the revised expression
\begin{equation}\label{new-guess}
    g_2
    =
    4\sqrt{g_4} + 12 g_r
    + \frac{16 g_r^2}{\sqrt{g_4}}
      \lr{1 - \exp\lr{-\dfrac{3 \sqrt{g_4}}{4 g_r}}}
    - \frac{g_4}{2 g_r}\,
      \psi_1\!\lr{\frac{\sqrt{g_4}}{4 g_r}} \, ,
\end{equation}
which is shown in Figure~\ref{figure:13}.

As shown in Figures~\ref{figure:09} and \ref{figure:13}, the present data
do not allow us to distinguish between the original ansatz
\eqref{old-guess} and the modified expression \eqref{new-guess} in the
$g_4 \to 0$ regime; both describe the simulations equally well within
statistical uncertainties. The essential difference is that the
expression \eqref{new-guess} includes a genuinely nonperturbative
exponential dependence on $g_r$, which was absent in the earlier
proposal \eqref{old-guess}.

\color{black}
\acknowledgments

This research was funded by the Ministry of Science, Technological Development and Innovation, Republic of Serbia through grant agreements with the University of Belgrade – Faculty of Pharmacy (Nos. 451-03-65/2024-03/200161, 451-03-136/2025-03/200161, and 451-03-137/2025-03/200161), by Comenius University in Bratislava under grant No.  UK/1082/2025, and by the VEGA project VEGA-1/0025/23, Matrix Models and Quantum Gravity.

Additional support was provided by COST Action CaLISTA (CA21109, E-COST-GRANT-CA21109-20f70f73), the CEEPUS network RS-1514-04-2324 (Umbrella) – Quantum Spacetime, Gravitation and Cosmology (M-RS-1514-2324-181120, M-RS-1514-2324-181129, M-RS-1514-2425-197018), and the Hungarian Collegium Talentum Programme.

\bibliographystyle{JHEP}
\bibliography{bibliography}

@article{Prekrat:2020ptq,
    author = "Prekrat, Dragan and Todorovi\'c-Vasovi\'c, Kristina Neli and Rankovi\'c, Dragana",
    title = "{Detecting scaling in phase transitions on the truncated Heisenberg algebra}",
    eprint = "2002.05704",
    archivePrefix = "arXiv",
    primaryClass = "hep-th",
    doi = "https://doi.org/10.1007/JHEP03(2021)197",
    journal = "JHEP\phantom{}",
    volume = "03",
    pages = "197",
    year = "2021"
}

@article{PhysRev.71.38,
    title = {Quantized Space-Time},
    author = {Snyder, Hartland S.},
    journal = {Phys. Rev.},
    volume = {71},
    issue = {1},
    pages = {38--41},
    numpages = {0},
    year = {1947},
    month = {Jan},
    publisher = {American Physical Society},
    doi = {10.1103/PhysRev.71.38},
    url = {https://link.aps.org/doi/10.1103/PhysRev.71.38}
}

@article{Bietenholz:2014sza,
    author = "Bietenholz, Wolfgang and Hofheinz, Frank and Mej\'\i{}a-D\'\i{}az, H\'ector and Panero, Marco",
    editor = "Delepine, David and Napsuciale, Mauro and Ibarguen, Humberto Salazar",
    title = "{Scalar fields in a non-commutative space}",
    eprint = "1402.4420",
    archivePrefix = "arXiv",
    primaryClass = "hep-th",
    reportNumber = "IFT-UAM-CSIC-14-006",
    doi = "10.1088/1742-6596/651/1/012003",
    journal = "J. Phys. Conf. Ser.",
    volume = "651",
    number = "1",
    pages = "012003",
    year = "2015"
}

@article{Dolan:2001gn,
    author = "Dolan, Brian P. and O'Connor, Denjoe and Presnajder, P.",
    title = "{Matrix $\phi^4$ models on the fuzzy sphere and their continuum limits}",
    eprint = "hep-th/0109084",
    archivePrefix = "arXiv",
    doi = "10.1088/1126-6708/2002/03/013",
    journal = "JHEP\phantom{}",
    volume = "03",
    pages = "013",
    year = "2002"
}

@article{Grosse:2003nw,
    author = "Grosse, Harald and Wulkenhaar, Raimar",
    title = "{Renormalization of $\phi^4$ theory on noncommutative $\mathbb{R}^2$ in the matrix base}",
    eprint = "hep-th/0307017",
    archivePrefix = "arXiv",
    doi = "10.1088/1126-6708/2003/12/019",
    journal = "JHEP\phantom{}",
    volume = "12",
    pages = "019",
    year = "2003"
}

@article{Grosse:2004yu,
    author = "Grosse, Harald and Wulkenhaar, Raimar",
    title = "{Renormalization of $\phi^4$ theory on noncommutative R**4 in the matrix base}",
    eprint = "hep-th/0401128",
    archivePrefix = "arXiv",
    doi = "10.1007/s00220-004-1285-2",
    journal = "Commun. Math. Phys.",
    volume = "256",
    pages = "305--374",
    year = "2005"
}

@article{Buric:2009ss,
    author = "Buri\'{c}, Maja and Wohlgenannt, Michael",
    title = "{Geometry of the Grosse-Wulkenhaar Model}",
    eprint = "0902.3408",
    archivePrefix = "arXiv",
    primaryClass = "hep-th",
    reportNumber = "DISTA-UPO-08",
    doi = "10.1007/JHEP03(2010)053",
    journal = "JHEP\phantom{}",
    volume = "03",
    pages = "053",
    year = "2010"
}

@article{Gubser:2000cd,
    author = "Gubser, Steven S. and Sondhi, Shivaji L.",
    title = "{Phase structure of noncommutative scalar field theories}",
    eprint = "hep-th/0006119",
    archivePrefix = "arXiv",
    reportNumber = "PUPT-1936",
    doi = "10.1016/S0550-3213(01)00108-0",
    journal = "Nucl. Phys. B",
    volume = "605",
    pages = "395--424",
    year = "2001"
}

@article{Castorina:2007za,
    author = "Castorina, Paolo and Zappala, Dario",
    title = "{Spontaneous breaking of translational invariance in non-commutative lambda $\phi^4$ theory in two dimensions}",
    eprint = "0711.2659",
    archivePrefix = "arXiv",
    primaryClass = "hep-th",
    doi = "10.1103/PhysRevD.77.027703",
    journal = "Phys. Rev. D",
    volume = "77",
    pages = "027703",
    year = "2008"
}

@article{Mejia-Diaz:2014lza,
    author = "Mej\'\i{}a-D\'\i{}az, H\'ector and Bietenholz, Wolfgang and Panero, Marco",
    title = "{The continuum phase diagram of the 2d non-commutative $\lambda \phi^4$ model}",
    eprint = "1403.3318",
    archivePrefix = "arXiv",
    primaryClass = "hep-lat",
    reportNumber = "IFT-UAM-CSIC-14-014",
    doi = "10.1007/JHEP10(2014)056",
    journal = "JHEP\phantom{}",
    volume = "10",
    pages = "056",
    year = "2014"
}

@article{Buric:2016lly,
    author = "Buri\'c, Maja and Nenadovi\'c, Luka and Prekrat, Dragan",
    title = "{One-loop structure of the $U(1)$ gauge model on the truncated Heisenberg space}",
    eprint = "1610.01429",
    archivePrefix = "arXiv",
    primaryClass = "hep-th",
    doi = "10.1140/epjc/s10052-016-4522-x",
    journal = "Eur. Phys. J. C",
    volume = "76",
    number = "12",
    pages = "672",
    year = "2016"
}

@article{Tekel:2015uza,
    author = "Tekel, Juraj",
    title = "{Phase structure of fuzzy field theories and multitrace matrix models}",
    eprint = "1512.00689",
    archivePrefix = "arXiv",
    primaryClass = "hep-th",
    journal = "Acta Phys. Slov.",
    volume = "65",
    number = "5",
    pages = "369--468",
    year = "2015",
    url = "http://www.physics.sk/aps/pub.php?y=2015&pub=aps-15-05"
}

@book{Ydri:2015zba,
    author = "Ydri, Badis",
    title = "{Computational Physics}: {An Introduction to Monte Carlo Simulations of Matrix Field Theory}",
    eprint = "1506.02567",
    archivePrefix = "arXiv",
    primaryClass = "hep-lat",
    doi = "10.1142/10283",
    isbn = "978-981-320-021-0, 978-981-320-023-4",
    publisher = "World Scientific",
    address = "Singapore",
    year = "2017"
}

@article{Kovacik:2018thy,
    author = "Kov\'a\v{c}ik, Samuel and O'Connor, Denjoe",
    title = "{Triple Point of a Scalar Field Theory on a Fuzzy Sphere}",
    eprint = "1805.08111",
    archivePrefix = "arXiv",
    primaryClass = "hep-th",
    reportNumber = "DIAS-STP-18-02",
    doi = "10.1007/JHEP10(2018)010",
    journal = "JHEP\phantom{}",
    volume = "10",
    pages = "010",
    year = "2018"
}

@article{Subjakova:2020shi,
    author = "\v{S}ubjakov\'a, M\'aria and Tekel, Juraj",
    title = "{Second moment fuzzy-field-theory-like matrix models}",
    eprint = "2002.02317",
    archivePrefix = "arXiv",
    primaryClass = "hep-th",
    doi = "10.1007/JHEP06(2020)088",
    journal = "JHEP\phantom{}",
    volume = "06",
    pages = "088",
    year = "2020"
}

@article{Franchino-Vinas:2019nqy,
    author = "Franchino-Vi\~nas, Sebasti\'an A. and Mignemi, Salvatore",
    title = "{Asymptotic freedom for $\lambda \phi ^4_{\star }$ QFT in Snyder\textendash{}de Sitter space}",
    eprint = "1911.08921",
    archivePrefix = "arXiv",
    primaryClass = "hep-th",
    doi = "10.1140/epjc/s10052-020-7918-6",
    journal = "Eur. Phys. J. C",
    volume = "80",
    number = "5",
    pages = "382",
    year = "2020"
}

@article{Prekrat:2021uos,
    author = "Prekrat, Dragan",
    title = "{Renormalization footprints in the phase diagram of the Grosse-Wulkenhaar model}",
    eprint = "2104.00657",
    archivePrefix = "arXiv",
    primaryClass = "hep-th",
    doi = "10.1103/PhysRevD.104.114505",
    journal = "Phys. Rev. D",
    volume = "104",
    number = "11",
    pages = "114505",
    year = "2021"
}

@misc{betancourt2018conceptual,
      title={A Conceptual Introduction to Hamiltonian Monte Carlo}, 
      author={Michael Betancourt},
      year={2018},
      eprint={1701.02434},
      archivePrefix={arXiv},
      primaryClass={stat.ME}
}

@article{Vinas:2014exa,
    author = "Vi\~nas, Sebasti\'an Franchino and Pisani, Pablo",
    title = "{Worldline approach to the Grosse-Wulkenhaar model}",
    eprint = "1406.7336",
    archivePrefix = "arXiv",
    primaryClass = "hep-th",
    doi = "10.1007/JHEP11(2014)087",
    journal = "JHEP\phantom{}",
    volume = "11",
    pages = "087",
    year = "2014"
}

@PHDTHESIS{Wulkenhaar:habilitation2004,
   author       = "Wulkenhaar, Raimar", 
   title        = "{Renormalisation of noncommutative $\phi_4^4$-theory to all orders}", 
   school       = "TU Wien, Faculty of Physics", 
   year         = "2004", 
   type         = "{Habilitation} thesis", 
   address      = "", 
   month        = "", 
   note         = "https://ivv5hpp.uni-muenster.de/u/raimar/publications/habilitation/raimar-habil.pdf", 
   url          = "https://ivv5hpp.uni-muenster.de/u/raimar/publications/habilitation/raimar-habil.pdf",
}

@article{Kovacik:2022kfh,
    author = "Kov\'a\v{c}ik, Samuel and Tekel, Juraj",
    title = "{Eigenvalue-flipping algorithm for matrix Monte Carlo}",
    eprint = "2203.05422",
    archivePrefix = "arXiv",
    primaryClass = "hep-lat",
    doi = "10.1007/JHEP04(2022)149",
    journal = "JHEP\phantom{}",
    volume = "04",
    pages = "149",
    year = "2022"
}

@article{Subjakova:2020prh,
    author = "\v{S}ubjakov\'a, M\'aria and Tekel, Juraj",
    title = "{Fuzzy field theories and related matrix models}",
    eprint = "2006.12605",
    archivePrefix = "arXiv",
    primaryClass = "hep-th",
    doi = "10.22323/1.376.0189",
    journal = "PoS",
    volume = "CORFU2019",
    pages = "189",
    year = "2020"
}

@article{Kanomata:2022pdo,
    author = "Kanomata, Naoyuki and Sako, Akifumi",
    title = "{Exact solution of the $\Phi_{2}^{3}$ finite matrix model}",
    eprint = "2205.15798",
    archivePrefix = "arXiv",
    primaryClass = "hep-th",
    doi = "10.1016/j.nuclphysb.2022.115892",
    journal = {Nuclear Physics B},
    issn = {0550-3213},
    volume = {982},
    pages = {115892},
    month = "5",
    year = "2022"
}

@article{Jeevanesan:2019jty,
    author = "Jeevanesan, Bhilahari and Moroz, Sergej",
    title = "{Thermodynamics of two-dimensional bosons in the lowest Landau level}",
    eprint = "1910.07808",
    archivePrefix = "arXiv",
    primaryClass = "cond-mat.quant-gas",
    doi = "10.1103/PhysRevResearch.2.033323",
    journal = "Phys. Rev. Res.",
    volume = "2",
    number = "3",
    pages = "033323",
    year = "2020"
}

@article{Steinacker:2011ix,
    author = "Steinacker, Harold",
    editor = "Barrett, John and Giesel, Kristina and Hellmann, Frank and Jonke, Larisa and Krajewski, Thomas and Lewandowski, Jerzy and Rovelli, Carlo and Sahlmann, Hanno and Steinacker, Harold",
    title = "{Non-commutative geometry and matrix models}",
    eprint = "1109.5521",
    archivePrefix = "arXiv",
    primaryClass = "hep-th",
    doi = "10.22323/1.140.0004",
    journal = "PoS",
    volume = "QGQGS2011",
    pages = "004",
    year = "2011"
}

@misc{Eynard:2015aea,
    author = "Eynard, Bertrand and Kimura, Taro and Ribault, Sylvain",
    title = "{Random matrices}",
    eprint = "1510.04430",
    archivePrefix = "arXiv",
    primaryClass = "math-ph",
    month = "10",
    year = "2015"
}

@article{PhysRevLett.94.168103,
  title = {Enumeration of RNA Structures by Matrix Models},
  author = {Vernizzi, Graziano and Orland, Henri and Zee, A.},
  journal = {Phys. Rev. Lett.},
  volume = {94},
  issue = {16},
  pages = {168103},
  numpages = {4},
  year = {2005},
  month = {Apr},
  publisher = {American Physical Society},
  doi = {10.1103/PhysRevLett.94.168103},
  url = {https://link.aps.org/doi/10.1103/PhysRevLett.94.168103}
}

@article{Weidenmuller:2008vb,
    author = "Weidenmuller, H. A. and Mitchell, G. E.",
    title = "{Random Matrices and Chaos in Nuclear Physics. Part 1. Nuclear Structure}",
    eprint = "0807.1070",
    archivePrefix = "arXiv",
    primaryClass = "nucl-th",
    doi = "10.1103/RevModPhys.81.539",
    journal = "Rev. Mod. Phys.",
    volume = "81",
    pages = "539--589",
    year = "2009"
}

@article{Guhr:1997ve,
    author = "Guhr, Thomas and Muller-Groeling, Axel and Weidenmuller, Hans A.",
    title = "{Random matrix theories in quantum physics: Common concepts}",
    eprint = "cond-mat/9707301",
    archivePrefix = "arXiv",
    reportNumber = "H-V27-1997",
    doi = "10.1016/S0370-1573(97)00088-4",
    journal = "Phys. Rept.",
    volume = "299",
    pages = "189--425",
    year = "1998"
}

@book{Akemann:2011csh,
    author = "Akemann, Gernot and Baik, Jinho and Di Francesco, Philippe",
    title = "{The Oxford Handbook of Random Matrix Theory}",
    isbn = "978-0-19-957400-1",
    publisher = "Oxford University Press",
    series = "Oxford Handbooks in Mathematics",
    month = "9",
    year = "2011"
}

@article{Beenakker:2014zza,
    author = "Beenakker, C. W. J.",
    title = "{Random-matrix theory of Majorana fermions and topological superconductors}",
    eprint = "1407.2131",
    archivePrefix = "arXiv",
    primaryClass = "cond-mat.mes-hall",
    doi = "10.1103/RevModPhys.87.1037",
    journal = "Rev. Mod. Phys.",
    volume = "87",
    pages = "1037",
    year = "2015"
}

@article{Loll:2019rdj,
    author = "Loll, R.",
    title = "{Quantum Gravity from Causal Dynamical Triangulations: A Review}",
    eprint = "1905.08669",
    archivePrefix = "arXiv",
    primaryClass = "hep-th",
    doi = "10.1088/1361-6382/ab57c7",
    journal = "Class. Quant. Grav.",
    volume = "37",
    number = "1",
    pages = "013002",
    year = "2020"
}

@article{Prekrat:2022sir,
    author = "Prekrat, D. and Rankovi\'c, D. and Todorovi\'c-Vasovi\'c, N. K. and Kov\'a\v{c}ik, S. and Tekel, J.",
    title = "{Approximate treatment of noncommutative curvature in quartic matrix model}",
    eprint = "2209.00592",
    archivePrefix = "arXiv",
    primaryClass = "hep-th",
    doi = "10.1007/JHEP01(2023)109",
    journal = "JHEP",
    volume = "01",
    pages = "109",
    year = "2023"
}

@article{Kanomata:2023mni,
    author = "Kanomata, Naoyuki and Sako, Akifumi",
    title = "{Exact Solutions v.s. Perturbative Calculations of Finite $\Phi^{3}$-$\Phi^{4}$ Hybrid-Matrix-Model}",
    eprint = "2304.10364",
    archivePrefix = "arXiv",
    primaryClass = "hep-th",
    year = "2023"
}

@article{Hossenfelder:2012jw,
    author = "Hossenfelder, Sabine",
    title = "{Minimal Length Scale Scenarios for Quantum Gravity}",
    eprint = "1203.6191",
    archivePrefix = "arXiv",
    primaryClass = "gr-qc",
    doi = "10.12942/lrr-2013-2",
    journal = "Living Rev. Rel.",
    volume = "16",
    pages = "2",
    year = "2013"
}

@article{Franchino-Vinas:2021bcl,
    author = "Franchino-Vi\~nas, S. A. and Mignemi, S.",
    title = "{The Snyder-de Sitter scalar \ensuremath{\varphi}\ensuremath{\star}4 quantum field theory in D = 2}",
    eprint = "2104.00043",
    archivePrefix = "arXiv",
    primaryClass = "hep-th",
    doi = "10.1016/j.nuclphysb.2022.115871",
    journal = "Nucl. Phys. B",
    volume = "981",
    pages = "115871",
    year = "2022"
}

@article{Ambjorn:2002nj,
    author = "Ambjorn, J. and Catterall, S.",
    title = "{Stripes from (noncommutative) stars}",
    eprint = "hep-lat/0209106",
    archivePrefix = "arXiv",
    reportNumber = "SU-4252-767",
    doi = "10.1016/S0370-2693(02)02906-4",
    journal = "Phys. Lett. B",
    volume = "549",
    pages = "253--259",
    year = "2002"
}

@article{Rosaler:2021quv,
    author = "Rosaler, Joshua",
    title = "{Dogmas of Effective Field Theory: Scheme Dependence, Fundamental Parameters, and the Many Faces of the Higgs Naturalness Principle}",
    doi = "10.1007/s10701-021-00510-4",
    journal = "Found. Phys.",
    volume = "52",
    number = "1",
    pages = "2",
    year = "2022"
}

@article{Fukuda:2019pzs,
    author = {Fukuda, Motohisa and K\"onig, Robert and Nechita, Ion},
    title = "{RTNI\textemdash{}A symbolic integrator for Haar-random tensor networks}",
    eprint = "1902.08539",
    archivePrefix = "arXiv",
    primaryClass = "quant-ph",
    doi = "10.1088/1751-8121/ab434b",
    journal = "J. Phys. A",
    volume = "52",
    number = "42",
    pages = "425303",
    year = "2019"
}

@article{Fukuda:2023xmz,
    author = "Fukuda, Motohisa",
    title = "{Symbolically integrating tensor networks over various random tensors by the second version of Python RTNI}",
    eprint = "2309.01167",
    archivePrefix = "arXiv",
    primaryClass = "physics.comp-ph",
    month = "9",
    year = "2023"
}

@article{SZABO2003207,
title = {Quantum field theory on noncommutative spaces},
journal = {Physics Reports},
volume = {378},
number = {4},
pages = {207-299},
year = {2003},
issn = {0370-1573},
doi = {https://doi.org/10.1016/S0370-1573(03)00059-0},
url = {https://www.sciencedirect.com/science/article/pii/S0370157303000590},
author = {Richard J. Szabo}
}

@article{Prekrat:2023fts,
    author = "Prekrat, Dragan and Rankovi\'c, Dragana and Todorovi\'c-Vasovi\'c, Neli Kristina and Kov\'a\v{c}ik, Samuel and Tekel, Juraj",
    title = "{Phase transitions in a \ensuremath{\Phi}4 matrix model on a curved noncommutative space}",
    eprint = "2310.10794",
    archivePrefix = "arXiv",
    primaryClass = "hep-th",
    doi = "10.1142/S0217751X23430029",
    journal = "Int. J. Mod. Phys. A",
    volume = "38",
    number = "32",
    pages = "2343002",
    year = "2023"
}

@PHDTHESIS{Prekrat:2023thesis,
   author       = "Prekrat, Dragan", 
   title        = "Phase transitions in matrix models on the truncated Heisenberg space", 
   school       = "University of Belgrade, Faculty of Physics", 
   year         = "2023", 
   
   type         = "{Ph.D.} thesis", 
   address      = "", 
   month        = "03", 
   note         = "https://hdl.handle.net/21.15107/rcub_nardus_21600", 
   url          = "https://hdl.handle.net/21.15107/rcub_nardus_21600",
}

@inproceedings{Benedek:2023stconf,
  author       = {Bukor, Benedek and Tekel, Juraj},
  booktitle    = {Proceedings of the Student Science Conference 2023},
  title        = {Second order kinetic term effective actions for matrix model description of fuzzy field theories},
  eventdate    = {April 27, 2023},
  venue        = {Faculty of Mathematics, Physics and Informatics, Comenius University, Bratislava},
  isbn         = {978-80-8147-136-0},
  publisher    = {FMFI UK, Bratislava, https://zona.fmph.uniba.sk/fileadmin/fmfi/studentska{\_}vedecka{\_}konferencia/zbierka2023/ svk2023{\_}zbornik.pdf},
  date         = {2023}
}

@article{Prekrat:2024rmq,
    author = "Prekrat, D. and Rankovic, D. and Minic, M. and Todorovic-Vasovic, N. K. and Kov\'acik, S. and Tekel, J.",
    title = "{(Non)renormalizable noncommutativity in (non)uniform phase}",
    doi = "10.22323/1.463.0269",
    journal = "PoS",
    volume = "CORFU2023",
    pages = "269",
    year = "2024"
}

@article{10.1093/mnras/staf292,
    author = {Shamir, Lior},
    title = {The distribution of galaxy rotation in JWST Advanced Deep Extragalactic Survey},
    journal = {Monthly Notices of the Royal Astronomical Society},
    volume = {538},
    number = {1},
    pages = {76-91},
    year = {2025},
    month = {02},
    issn = {0035-8711},
    doi = {10.1093/mnras/staf292},
    url = {https://doi.org/10.1093/mnras/staf292},
    eprint = {https://academic.oup.com/mnras/article-pdf/538/1/76/61934633/staf292.pdf},
}

@article{Karabali:2002im,
    author = "Karabali, Dimitra and Nair, V. P.",
    title = "{Quantum Hall effect in higher dimensions}",
    eprint = "hep-th/0203264",
    archivePrefix = "arXiv",
    reportNumber = "CCNY-HEP-02-03",
    doi = "10.1016/S0550-3213(02)00634-X",
    journal = "Nucl. Phys. B",
    volume = "641",
    pages = "533--546",
    year = "2002"
}

@article{Karabali:2006eg,
    author = "Karabali, Dimitra and Nair, V. P.",
    title = "{Quantum Hall effect in higher dimensions, matrix models and fuzzy geometry}",
    eprint = "hep-th/0606161",
    archivePrefix = "arXiv",
    reportNumber = "CCNY-HEP06-8",
    doi = "10.1088/0305-4470/39/41/S05",
    journal = "J. Phys. A",
    volume = "39",
    pages = "12735--12764",
    year = "2006"
}

@article{Subjakova:2020gdh,
    author = "\v{S}ubjakov\'a, M\'aria and Tekel, Juraj",
    title = "{Beyond second-moment approximation in fuzzy-field-theory-like matrix models}",
    eprint = "2109.03363",
    archivePrefix = "arXiv",
    primaryClass = "hep-th",
    doi = "10.1007/JHEP02(2022)065",
    journal = "JHEP",
    volume = "22",
    pages = "065",
    year = "2020"
}

@article{Ydri:2021cam,
    author = "Ydri, Badis and Khaled, Ramda and Soudani, Cherine",
    title = "{Quantized noncommutative geometry from multitrace matrix models}",
    eprint = "2110.06677",
    archivePrefix = "arXiv",
    primaryClass = "hep-th",
    doi = "10.1142/S0217751X2250052X",
    journal = "Int. J. Mod. Phys. A",
    volume = "37",
    number = "10",
    pages = "2250052",
    year = "2022"
}

@article{Ydri:2020efr,
    author = "Ydri, Badis and Ahmim, Rachid",
    title = "{Wilsonian renormalization group for a multitrace matrix model}",
    eprint = "2008.09564",
    archivePrefix = "arXiv",
    primaryClass = "hep-th",
    doi = "10.1142/S0217751X22501652",
    journal = "Int. J. Mod. Phys. A",
    volume = "37",
    number = "27",
    pages = "2250165",
    year = "2022"
}

@article{Kovacik:2020cod,
    author = "Kov\'a\v{c}ik, Samuel and O'Connor, Denjoe and Asano, Yuhma",
    title = "{The nonperturbative phase diagram of the bosonic BMN matrix model}",
    eprint = "2004.05820",
    archivePrefix = "arXiv",
    primaryClass = "hep-th",
    doi = "10.22323/1.376.0221",
    journal = "PoS",
    volume = "CORFU2019",
    pages = "221",
    year = "2020"
}

@article{Kovacik:2023zab,
    author = "Kov\'a\v{c}ik, Samuel and Tekel, Juraj",
    title = "{Fuzzy onionlike space as a matrix model}",
    eprint = "2309.00576",
    archivePrefix = "arXiv",
    primaryClass = "hep-th",
    doi = "10.1103/PhysRevD.109.105004",
    journal = "Phys. Rev. D",
    volume = "109",
    number = "10",
    pages = "105004",
    year = "2024"
}

@article{Bukor_2025,
    doi = {10.1088/1751-8121/addeac},
    url = {https://dx.doi.org/10.1088/1751-8121/addeac},
    year = {2025},
    month = {jun},
    publisher = {IOP Publishing},
    volume = {58},
    number = {25},
    pages = {255203},
    author = {Bukor, Benedek and Tekel, Juraj},
    title = {Cubic asymmetric multitrace matrix model},
    journal = {Journal of Physics A: Mathematical and Theoretical}
}

@article{Han:2019wue,
    author = "Han, Xizhi and Hartnoll, Sean A.",
    title = "{Deep Quantum Geometry of Matrices}",
    eprint = "1906.08781",
    archivePrefix = "arXiv",
    primaryClass = "hep-th",
    doi = "10.1103/PhysRevX.10.011069",
    journal = "Phys. Rev. X",
    volume = "10",
    number = "1",
    pages = "011069",
    year = "2020"
}

@article{Chen:2017kfj,
    author = "Chen, Hong Zhe and Karczmarek, Joanna L.",
    title = "{Entanglement entropy on a fuzzy sphere with a UV cutoff}",
    eprint = "1712.09464",
    archivePrefix = "arXiv",
    primaryClass = "hep-th",
    doi = "10.1007/JHEP08(2018)154",
    journal = "JHEP",
    volume = "08",
    pages = "154",
    year = "2018"
}

@article{Prekrat:2025ozs,
    author = "Prekrat, Dragan",
    title = "{Comment on ''Geometry of the Grosse-Wulkenhaar model''}",
    eprint = "2505.18123",
    archivePrefix = "arXiv",
    primaryClass = "hep-th",
    month = "5",
    year = "2025"
}

@article{Hessam:2021byc,
    author = "Hessam, Hamed and Khalkhali, Masoud and Pagliaroli, Nathan",
    title = "{Bootstrapping Dirac ensembles}",
    eprint = "2107.10333",
    archivePrefix = "arXiv",
    primaryClass = "hep-th",
    doi = "10.1088/1751-8121/ac5216",
    journal = "J. Phys. A",
    volume = "55",
    number = "33",
    pages = "335204",
    year = "2022"
}

@article{Hessam:2022gaw,
    author = "Hessam, Hamed and Khalkhali, Masoud and Pagliaroli, Nathan and Verhoeven, Luuk S.",
    title = "{From noncommutative geometry to random matrix theory}",
    eprint = "2204.14216",
    archivePrefix = "arXiv",
    primaryClass = "hep-th",
    doi = "10.1088/1751-8121/ac8fc5",
    journal = "J. Phys. A",
    volume = "55",
    number = "41",
    pages = "413002",
    year = "2022"
}

@article{Khalkhali:2023onm,
    author = "Khalkhali, Masoud and Pagliaroli, Nathan",
    title = "{Coloured combinatorial maps and quartic bi-tracial 2-matrix ensembles from noncommutative geometry}",
    eprint = "2312.10530",
    archivePrefix = "arXiv",
    primaryClass = "math-ph",
    doi = "10.1007/JHEP05(2024)186",
    journal = "JHEP",
    volume = "05",
    pages = "186",
    year = "2024"
}

@article{Khalkhali:2020djp,
    author = "Khalkhali, Masoud and Pagliaroli, Nathan",
    title = "{Phase Transition in Random Noncommutative Geometries}",
    eprint = "2006.02891",
    archivePrefix = "arXiv",
    primaryClass = "math-ph",
    doi = "10.1088/1751-8121/abd190",
    journal = "J. Phys. A",
    volume = "54",
    number = "3",
    pages = "035202",
    year = "2021"
}

@article{Barrett:2015foa,
    author = "Barrett, John W. and Glaser, Lisa",
    title = "{Monte Carlo simulations of random non-commutative geometries}",
    eprint = "1510.01377",
    archivePrefix = "arXiv",
    primaryClass = "gr-qc",
    doi = "10.1088/1751-8113/49/24/245001",
    journal = "J. Phys. A",
    volume = "49",
    number = "24",
    pages = "245001",
    year = "2016"
}

@article{Han:2017htd,
    author = "Han, J. L.",
    title = "{Observing Interstellar and Intergalactic Magnetic Fields}",
    doi = "10.1146/annurev-astro-091916-055221",
    journal = "Ann. Rev. Astron. Astrophys.",
    volume = "55",
    number = "1",
    pages = "111--157",
    year = "2017"
}

\end{document}